\documentclass[showpacs,amsmath,amssymb,twocolumn,pra,superscriptaddress,notitlepage]{revtex4-1}

\usepackage{qcircuit}
\usepackage{amsmath,bm}
\usepackage[dvips]{graphicx}
\usepackage{amsmath,amssymb,amsthm,mathrsfs,amsfonts,dsfont}
\usepackage{braket}
\usepackage{physics}
\usepackage{bm}
\usepackage{enumerate}
\usepackage{color}
\usepackage{graphicx}
\usepackage{appendix}
\usepackage{algorithm}
\usepackage{algorithmic}
\usepackage[justification=RaggedRight]{caption}
\usepackage[subrefformat=parens]{subcaption}
\usepackage{tabularx}
\newcommand{\YM}[1]{\textcolor{black}{#1}}
\newcommand{\YS}[1]{\textcolor{black}{#1}}

\newcolumntype{Y}{&gt;{\centering\arraybackslash}X} %中央揃え

\begin{document}

% --------------------  TITLE  --------------------

\title{
Quantum annealing with error mitigation
%intermediate-scale quantum computing with security inbuilt
}

% ------------  AUTHORS AND AFFILIATIONS ----------
\author{Yuta Shingu}
\email{shingu.yuta@aist.go.jp}
\affiliation{Department  of  Physics, Graduate School of Science, Tokyo  University  of  Science,  Shinjuku,  Tokyo  162-8601,  Japan.}
\affiliation{Research Center for Emerging Computing Technologies (RCECT),  National  Institute  of  Advanced  Industrial  Science  and  Technology  (AIST),1-1-1  Umezono,  Tsukuba,  Ibaraki  305-8568,  Japan.}

\author{Tetsuro Nikuni}\email{nikuni@rs.tus.ac.jp}
\affiliation{Department  of  Physics, Graduate School of Science,Tokyo  University  of  Science,  Shinjuku,  Tokyo  162-8601,  Japan.}

 \author{Shiro Kawabata}
 \affiliation{Research Center for Emerging Computing Technologies (RCECT),  National  Institute  of  Advanced  Industrial  Science  and  Technology  (AIST),1-1-1  Umezono,  Tsukuba,  Ibaraki  305-8568,  Japan.}
 \affiliation{NEC-AIST Quantum Technology Cooperative Research Laboratory, National Institute of Advanced Industrial Science and Technology (AIST), Tsukuba, Ibaraki 305-8568, Japan}

\author{Yuichiro Matsuzaki 
}
\email{matsuzaki.yuichiro@aist.go.jp}  
\affiliation{Research Center for Emerging Computing Technologies (RCECT),  National  Institute  of  Advanced  Industrial  Science  and  Technology  (AIST),1-1-1  Umezono,  Tsukuba,  Ibaraki  305-8568,  Japan.}
\affiliation{NEC-AIST Quantum Technology Cooperative Research Laboratory, National Institute of Advanced Industrial Science and Technology (AIST), Tsukuba, Ibaraki 305-8568, Japan}

% --------------------  ABSTRACT  --------------------

\begin{abstract}    
Quantum annealing (QA) is one of the efficient methods to calculate the ground-state energy of a problem Hamiltonian. In the absence of noise, QA can accurately estimate the ground-state energy if the adiabatic condition is satisfied. However, in actual physical implementation, systems suffer from decoherence. 
On the other hand, much effort has been paid into the noisy intermediate-scale quantum (NISQ) computation research. For practical NISQ computation, many error mitigation (EM) methods have been devised to remove noise effects.
In this paper, we propose a QA strategy combined with the EM method called dual-state purification to suppress the effects of decoherence. Our protocol consists of four parts; the conventional dynamics, single-qubit projective measurements, Hamiltonian dynamics corresponding to an inverse map of the first dynamics, and post-processing of measurement results.
Importantly, our protocol works without two-qubit gates, and so our protocol is suitable for the devices designed for practical QA. We also provide numerical calculations to show that our protocol leads to a more accurate estimation of the ground energy than the conventional QA under decoherence.
\end{abstract}

\maketitle

\section{Introduction}
Quantum annealing (QA)~\cite{kadowaki1998quantum,farhi2000quantum,farhi2001quantum} is a promising 
% \YMdel{quantum architectures} 
way to obtain a ground state \YM{of a problem Hamiltonian}.
% \YMdel{and a ground-state energy of a given problem Hamiltonian. }
\YM{Initially, the system is prepared as a ground state of a driver Hamiltonian.}
In QA, we adopt a time-dependent total Hamiltonian that changes from the driver Hamiltonian to the problem Hamiltonian, and we let the state evolve by such a Hamiltonian.
%After preparing a ground state of the driver Hamiltonian, we let the state evolve by the total Hamiltonian.
% \YMdel{
% To obtain them with QA, after one prepares a ground state of a transverse field which is called a driving Hamiltonian, the quantum state evolves with a time-dependent Hamiltonian which changes from the transverse field to the problem Hamiltonian.} \YMdel{If}
\YM{As long as}
the adiabatic condition is satisfied, 
% \YMdel{and the Hamiltonian can lead to the transitions between the initial state and the target state, 
% one can generate}
\YM{we obtain}
the ground state of the problem Hamiltonian \YM{after the dynamics without noise~\cite{paul1916adiabatische,kato1950adiabatic,amin2009consistency,dodin2021generalized}}. 
% \YMdel{owing to the dynamics. }

\YS{
% \YMdel{When the problem Hamiltonian is the Ising type, which has only diagonal terms,} 
%\YM{we can increase the success probability} to 
\YM{We can}
obtain 
the ground state of the \YM{Ising-type}
problem Hamiltonian 
% \YMdel{can be obtained after QA} 
by measuring the state \YM{after QA}
in the computational basis, \YM{as long as the state has a finite population of the ground state}~\cite{lechner2015quantum,kumar2018quantum,choi2011minor}. 
The 
% \YMdel{time-evolved} 
density matrix \YM{after the QA}
% \YMdel{is written as}
\YM{can be expanded by the energy eigenbasis as}
% \YMdel{$\rho=\sum_j p_j\ket{E_j}\bra{E_j}$,}
\YM{$\rho=\sum_{j,j'} d_{j,j'}\ket{E_j}\bra{E_{j'}}$}
where $d_{j,j'}$ denotes a coefficient,
$p_{j}\equiv d_{j,j}$ denotes a population, and $\ket{E_j}$ denotes an eigenvector of the problem Hamiltonian. In this case, the probability to obtain the ground state 
% \YMdel{by}
\YM{with}
the measurement in the computational basis is described 
% \YMdel{as}
\YM{by} $1-(1-p_0)^{N_{\rm{trial}}}$ where $p_0$ denotes the population of the ground state and $N_{\rm{trial}}$ denotes the number of trials.
\YM{Thus, we can obtain the ground state after many trials as long as $p_0$ has a non-zero value.}
% \YMdel{Thus, this probability increases in line with the number of trials if the ground-state population is finite.}
}

\YS{On the other hand, when the problem Hamiltonian contains non-diagonal elements,  
% \YMdel{the ground state of the problem Hamiltonian cannot be
% estimated
% by the measurements in the computational basis. }
\YM{the main aim is to estimate the energy of the ground state, which is the focus of our paper.
For such a Hamiltonian, we cannot obtain the ground state of the problem Hamiltonian
by the measurements in the computational basis after QA.
The expectation value of the problem Hamiltonian with the density matrix $\rho$ after QA is 
% \YMdel{defined}
\YM{given}
as $\ev{H_{\rm{P}}}=\mathrm{Tr}[H_{\rm{P}}\rho]=\sum_j p_j E_j$.
The problem Hamiltonian $H_{\mathrm{P}}$}
% \YMdel{Let us assume that} 
% \YMdel{is the summation of}
\YM{can be expanded by}
the products of the Pauli matrices 
% \YMdel{, like}
\YM{such as} $H_{\mathrm{P}}=\sum_i c_i \sigma_i$ where $c_i$ denotes a coefficient and $\sigma_i$ denotes a Pauli product.
% \YMdel{In this case, we 
% consider
% \YM{perform}
% %the 
% Pauli measurements to estimate the expected value of the energy.}  
To obtain the expectation value of the problem Hamiltonian, one needs to know the expectation value of each term by the measurements in the Pauli basis after QA and take the summation of  
% \YMdel{all the terms} 
\YM{these expectation values.}
% \YMdel{The expected value of the problem Hamiltonian with the density matrix $\rho$ after QA is defined as $\ev{H}=\mathrm{Tr}[H\rho]=\sum_j p_j E_j$, then $\ev{H}$ is obtained by performing a quite number of measurements.} 
% \YMdel{It means that if the populations of the excited states are not 0, the energy measured experimentally is not equal to the ground-state energy. Therefore, estimating the ground-state energy with high accuracy requires to generate a density matrix close to the ground state.}
\YM{Notably, a finite population of the excited state after QA leads to an error in the estimation of the ground-state energy. Therefore, for accurate estimation of the ground-state energy, we need to prepare the ground state with high fidelity.}
}

\begin{figure*}[htbp]
  \includegraphics[width=1\textwidth]{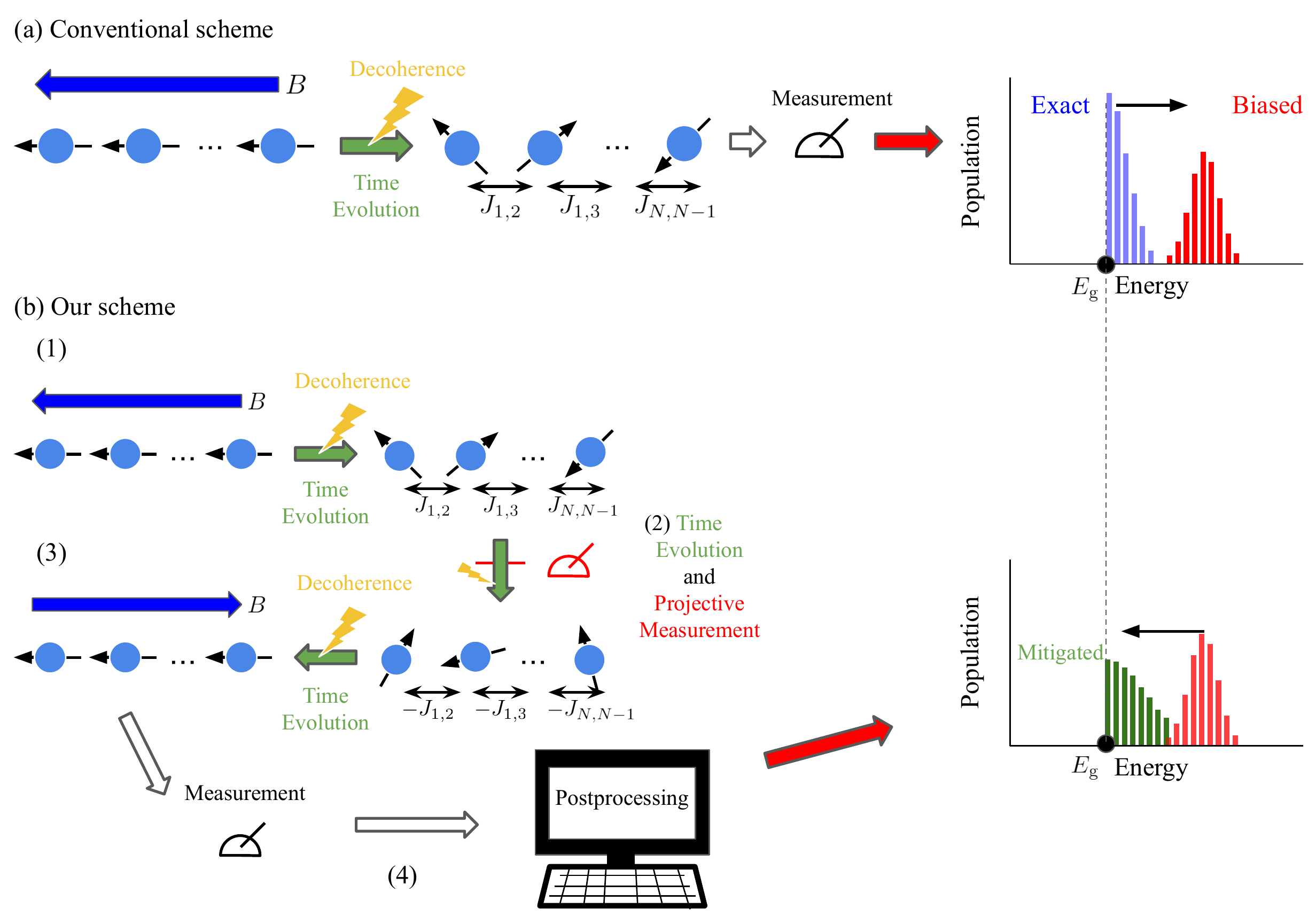}
  \caption{The conceptual diagram to indicate the difference between the conventional and our schemes in QA. (a)
  With the conventional scheme, 
  % \YMdel{in Fig.~\ref{fig: conceptual diagram} (a),} 
  one \YM{gradually} changes the total Hamiltonian from a driver Hamiltonian $H_{\mathrm{D}}$ to a problem Hamiltonian $H_{\mathrm{P}}$. 
  \textcolor{black}{The blue bars in the graph draw the exact populations without decoherence which can include the effects of non-adiabatic transitions. By decoherence, we actually obtain the biased populations as shown by the red bars.
  }
  In this case,
  the estimation of the ground-state energy $E_{\mathrm{g}}$ of $H_{\mathrm{P}}$ should be inaccurate if there is decoherence. 
  % \YMdel{On the other hand,}
  (b) With our scheme to incorporate the EM method, there are four parts:
  (1) the conventional dynamics, \textcolor{black}{(2) switching the sign of the coefficients in $H_{\mathrm{P}}$ and} single-qubit projective measurements,
  (3) Hamiltonian dynamics corresponding to an inverse map of the 
  % \YMdel{conventional QA}
  first dynamics, and (4) post-processing the measurement results.
  \textcolor{black}{The green bars illustrate the populations in which noise effects are mitigated.}
  % \YMdel{
  % we provide a QA schedule to incorporate the error mitigation method called dual-state purification} \YMdel{ as shown in Fig.~\ref{fig: conceptual diagram} (b).}
  % \YMdel{First, we perform the time evolution which is the same as the conventional scheme provides (see Fig.\ref{fig: conceptual diagram} (b)(1)). Then, we switch the sign of $H_{\mathrm{P}}$ (see Fig.\ref{fig: conceptual diagram} (b)(2)). In the middle of this switching, we implement the projective measurement of $H_{\mathrm{P}}$. Next, we again change the total Hamiltonian from $H_{\mathrm{P}}$ to $H_{\mathrm{D}}$ as shown in Fig.\ref{fig: conceptual diagram} (b)(3). After measurement and postprocessing as shown in Fig.\ref{fig: conceptual diagram} (b)(4), we obtain the expected value of $H_{\mathrm{P}}$ in which the effect of decoherence is mitigated.}
  \YM{Due to the EM method, we can estimate the ground-state energy more accurate than the conventional QA in Fig.~\ref{fig: conceptual diagram} (a).}
  }
    \label{fig: conceptual diagram}
\end{figure*}

\YM{However, QA suffers from non-adiabatic transitions and decoherence~\cite{morita2008mathematical,albert1961quantum,albert1962quantum,roland2005noise,aaberg2005quantum,albash2015decoherence,childs2001robustness,sarandy2005adiabatic}. If the dynamics are not slow enough to satisfy the adiabatic condition,  unwanted transitions from the ground state to the excited states occur. On the other hand, if the time scale of QA is comparable with or larger than the coherence time of the qubit, the system is affected by decoherence, which also induces unwanted transitions to the excited states. Thus, the estimation of the ground energy would be biased, as shown in Fig.~\ref{fig: conceptual diagram} (a).}
% \YMdel{
% Having said that, one does not always obtain the ground state though the above conditions are satisfied. The quantum state can be excited during the time evolution because of the decoherence derived from the environment. Therefore, many QA schemes have suffered from the effect of the environment.
% }

Meanwhile, many research works have been devoted theoretically and experimentally to realizing practical noisy intermediate-scale quantum (NISQ) computing~\cite{preskill2018quantum,endo2021hybrid,cerezo2021variational}.
We can use NISQ devices to perform quantum computation using tens to thousands of qubits, with $10^{-3}$ or lower gate errors.
% \YMdel{In many NISQ algorithms with variational quantum circuits that have been proposed to implement eigensolvers or simulations,}
\YM{Many algorithms for NISQ computing have been proposed.
Here, variational quantum circuits are typically used to generate a trial wave function to minimize a cost function~\cite{peruzzo2014variational,li2017efficient,mcclean2016theory,yuan2019theory,endo2018variational}.}
% \YMdel{it is an essential task to evaluate the expected values of some observables on the qubits.}
\YM{In these algorithms,
one needs to measure observables on the qubits corresponding to the trial wave function.}
% \YMdel{
% Though this estimation is usually prevented from obtaining the exact expected values because of the noise, one can suppress the effect of the noise owing to error mitigation methods. }
\YM{In the actual devices, noise prevents one from obtaining the accurate expectation values of the observable, degrading the algorithm performance.
Fortunately, sophisticated techniques called ``error mitigation" (EM) suppress the effect of noise by implementing additional quantum gates and post-processing with classical computation~\cite{endo2021hybrid,li2017efficient,kandala2019error,temme2017error,endo2018practical,song2019quantum,zhang2020error,mcardle2019error,bonet2018low,sun2021mitigating,larose2022mitiq,google2020hartree,mcclean2017hybrid,merkel2013self,stark2014self,greenbaum2015introduction,blume2017demonstration,strikis2021learning,czarnik2021error,wang2021scalable,o2021error,yoshioka2022generalized,cai2022quantum,cao2022mitigating}.
}
% \YMdel{Error mitigation is the means to modify the measurement result including the noise obtained with NISQ devices}
% \YMdel{circuits, preparing some ancilla qubits, or post-processing with classical computation.}

\YS{There are many types of EM. 
% \YMdel{It is known that}
The quasi-probability method is one way
% \YMdel{most efficient} 
to 
% \YMdel{remove}
\YM{mitigate the environmental effect~\cite{temme2017error,endo2018practical,huo2021self,takagi2021optimal,song2019quantum,zhang2020error,hakoshima2021relationship,sun2021mitigating,suzuki2022quantum}.}
% \YMdel{the bias derived from the environment. To obtain the computational result of the ideal quantum operation without the effect of the noise, one can represent the inverse map of each noise by inserting stochastic operations in the quantum circuit.}
\YM{By applying stochastic operations during the implementation of the quantum algorithm, one can effectively cancel out the environmental noise.}
However,
% \YMdel{since} 
this method requires 
\YM{accurate information about the noise model.}
% \YMdel{knowing the detail of the noise model, for example with the gate set tomography, }
\YM{If the environmental noise model does not change during the experiments, a so-called gate set tomography allows us to know the noise model with}
% \YMdel{this preprocessing leads to} 
the considerable measurement cost.
\YM{If the noise model changes in time, the quasi-probability methods cannot mitigate noise effects.}}

\YS{
On the other hand, the virtual distillation (VD) method, which is also known as the exponential error suppression (EES) method,
\YM{can mitigate the error without knowing the detail of the noise model~\cite{koczor2021exponential,huggins2021virtual,czarnik2021qubit,yamamoto2021error}.}
% \YMdel{avoids this problem effectively, i.e., one does not have to get the information of the noise before implementing the error mitigation method.}
Suppose that $N$ qubits are required to implement a quantum algorithm without EM.
In VD/EES, one needs to prepare $M$ copies of noisy quantum states $\rho$ composed of $N$ qubits to mitigate the error. We assume that these states are generated by the same quantum circuit  and influenced by the same noise model. 
% \YMdel{First, one prepares $M$ copies of noisy quantum states $\rho$ on $M$ different sets of $N$ qubits which are generated by the same quantum-circuit design and influenced by the same noise model. Then, the error-mitigated expected value of an observable}
\YM{Via the implementation of entangling gates between $M$ copies of a quantum state $\rho$, one can obtain}
%$O$ is computed as 
$\ev{O}^{(M)}_{\mathrm{VD}}=\mathrm{Tr}[O \rho^{(M)}_{\mathrm{VD}}]$ with $\rho^{(M)}_{\mathrm{VD}}=\rho^M / \mathrm{Tr}[\rho^M]$.
The advantage of this scheme is that the population of the dominant eigenvector of $\rho$ approaches unity, and so stochastic errors are exponentially suppressed as one increases $M$.
% \YMdel{Though one can suppress the effect of the stochastic errors, the coherent errors are intractable for this method. In addition, the number of the qubits needed by this method becomes $M$ times compared with the one that the original quantum computation requires without the error mitigation method as we mentioned above.}
\YM{However, VD/EES requires $MN$ qubits of the quantum state, and this is costly for NISQ
computers.}
}

% Notably, one error mitigation method has recently been proposed called dual-state purification. 
% \YMdel{Now, the recent error mitigation scheme called "dual-state purification" can overcome this problem. }
\YM{Recently, to overcome this problem, an alternative scheme called ``dual-state purification" was proposed~\cite{huo2022dual}.}
Let us consider the specific case $M=2$. 
% \YMdel{This scheme can prepare}
\YM{In this scheme, one can effectively prepare}
two copies of the quantum state on the same qubits. 
%One of 
%them can be 
% \YMdel{represented} 
\YM{one can
physically generate one of them} in an original quantum circuit, while 
% \YMdel{we can effectively prepare} 
the other \YM{one can be virtually prepared}
% \YMdel{copy} 
by the use of an inverse of the original circuit,
\YM{which is designed to be the
conjugate transpose of the original circuit
when there is no error.}
% \YMdel{Hence, the number of the qubits needed by this method is same as the the one the original algorithm requires without error mitigation although the circuit depth can become more than twice.}
% \YMdel{The inverse circuit is designed to be the conjugate transpose of the original circuit. when there is no error.}
By decomposing an observable $O$ into Pauli products $\{\sigma_i\}$, one obtains the expectation value $\ev{O}$ as $\Sigma a_i\ev{\sigma_i}$ where $a_i$ is a coefficient. In dual-state purification, one calculates each expectation value $\ev{\sigma_i}$ with each corresponding circuit and then takes the summation to compute the expectation value $\ev{O}$.
The entire circuit \YM{for dual-state purification}
is composed of the original circuit, the projective measurement of a Pauli product $\sigma_i$, the inverse circuit, \YM{and the projective measurement to the initial state}.
% \YMdel{Then, one needs to measure the state to observe how population the initial state is in the state. }  
% \YMdel{After postprocessing} 
After post-processing operations
with classical computation, one obtains the expectation value of $O$,
\YM{where stochastic errors during the implementation of quantum circuits are mitigated. Importantly, this scheme requires only $N$ qubits.}
% \YMdel{where the error is mitigated.}
% There is a naive question of whether one can mitigate the error in quantum annealing with error mitigation. 

\YM{
In this paper, we propose QA with 
dual-state purification.
% \YMdel{the error mitigation method.}
With naive applications of dual-state purification to QA, the use of the so-called reverse quantum annealing seems to be promising for the construction of the inverse dynamics of QA.
More specifically, the conventional QA is firstly performed where we gradually decrease (increase) the driver (problem) Hamiltonian,
 projective measurements are performed,
and then the reverse QA is implemented where
we gradually increase (decrease) the driver (problem) Hamiltonian. However, we show that this strategy could fail to estimate the ground-state energy, and
dual-state purification provides an unphysical density matrix with negative eigenvalues.
To overcome this problem, we propose modified scheduling to apply dual-state purification to QA. 
% In our scheduling, 
As shown in Fig.~\ref{fig: conceptual diagram} (b),
(1) we perform the same dynamics as the conventional QA, (2) we change the sign of the problem Hamiltonian in an adiabatic way and perform single-qubit projective measurements, (3) we gradually decrease (increase) the driver (problem) Hamiltonian, and (4) we post-process measurement results. In this case, we show that the density matrix produced by dual-state purification does not have negative eigenvalues. 
% (positive semi definitive density matrix???)
Moreover, we numerically demonstrate that our strategy provides a more accurate estimation than the conventional QA.}

% \YMdel{We prepare the copies by executing an annealing schedule; i.e., the quantum state in the annealing device evolves with the original quantum annealing, then it is measured by the projective measurement to obtain the expected value of the problem Hamiltonian, and finally it evolves with the inverse dynamics and is measured in the basis including the initial state.}

% \YMdel{suggest quantum annealing incorporating the error mitigation method that has recently been proposed.}
% We mitigate the error in quantum annealing by the inverse dynamics
% \YM{Let us define $U$ as the ideal }

% \YMdel{requires the inverse map of the original operation to prepare two copies of the quantum state on the same qubits. The inverse map should be designed as it is the conjugate transpose of the original operation when there is no error. We prepare the copies by executing an annealing schedule; i.e., the quantum state in the annealing device evolves with the original quantum annealing, then it is measured by the projective measurement to obtain the expected value of the problem Hamiltonian, and finally it evolves with the inverse dynamics and is measured with the initial state. The advantage of our scheme is that we can suppress the error regardless of the type of the problem Hamiltonian. In addition, our scheme is feasible even if the error model is unknown.}

\section{Quantum annealing}
\begin{figure}[htbp]
  \includegraphics[width=0.3\textwidth]{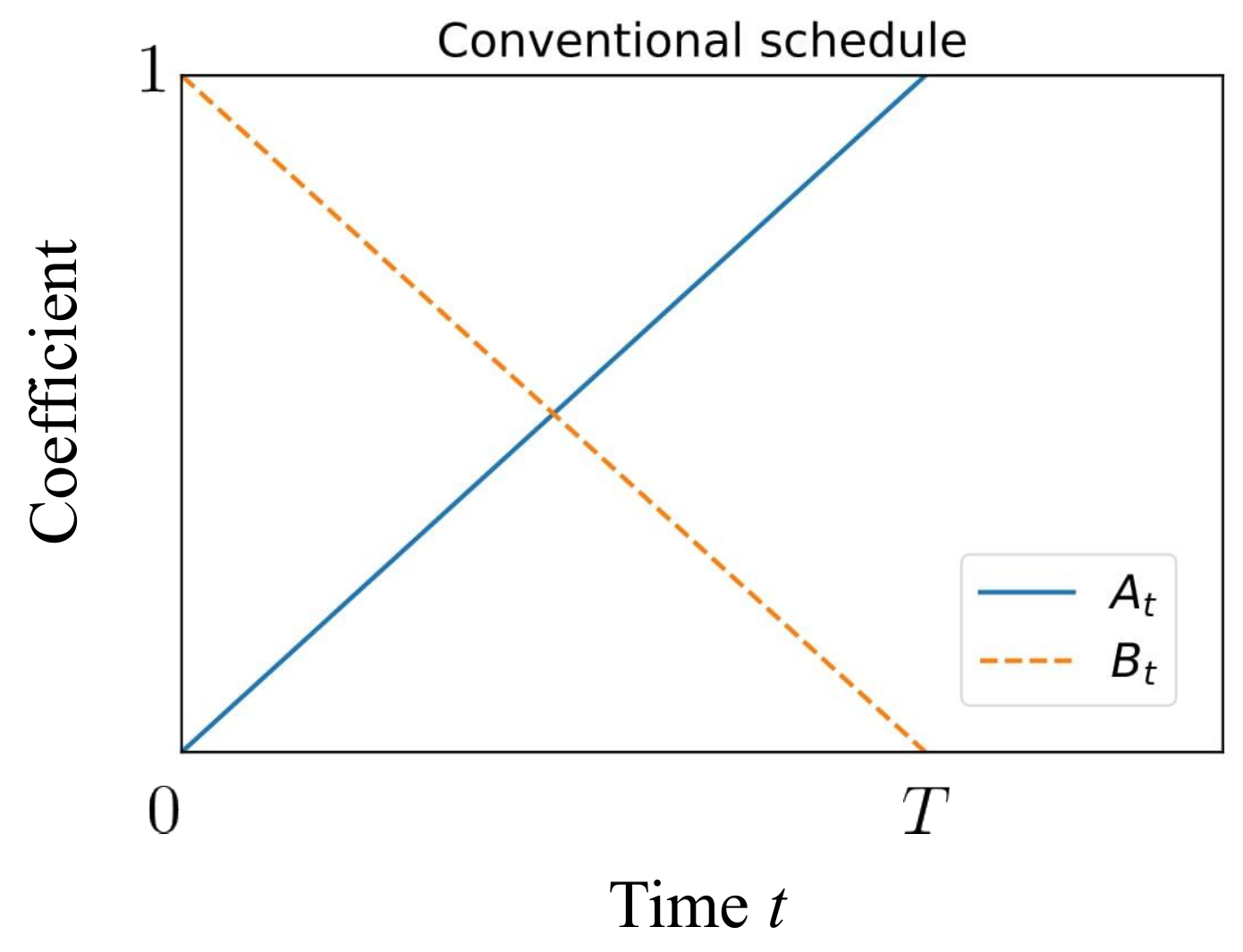}
  \caption{The 
  % \YMdel{conventional} 
  schedule of
  \YM{the conventional}
  QA. $A_t$ ($B_t$) represents the coefficient of a problem Hamiltonian $H_{\mathrm{P}}$ (a
  % \YMdel{ transverse field}
  \YM{driver}
  Hamiltonian $H_{\mathrm{D}}$). 
  % \YMdel{One}
  \YM{We}
  linearly change $A_t$ ($B_t$) from $0$  to $1$ (from $1$ to $0$) for an annealing time $T$.}
    \label{fig: conventional annealing schedule}
\end{figure}

\YS{Let us review QA to obtain the ground state and the ground-state energy of a problem Hamiltonian $H_{\mathrm{P}}$~\cite{kadowaki1998quantum,farhi2000quantum,farhi2001quantum}. 
We choose the independent spin model with the transverse field (i.e., $H_{\mathrm{D}}=-B\sum^N_{i=1}\hat{\sigma}^x_i$) as a driver Hamiltonian where $N$ denotes the number of qubits and $B$ denotes a coefficient.
Also, the system is initially prepared as 
$\ket{+}^{\otimes N}$
which is the ground state of the driver Hamiltonian $H_{\mathrm{D}}$ where $\ket{+}=\frac{1}{\sqrt{2}}(\ket{0}+\ket{1})$ denotes the eigenstate of $\hat{\sigma}^x$ and $\ket{0}$ and $\ket{1}$ denote the eigenstates of $\hat{\sigma}^z$.
We \YM{change}
% \YMdel{consider changing} 
the total Hamiltonian $H(t)$ from the driver Hamiltonian $H_{\mathrm{D}}$ 
% \YMdel{which we choose as the transverse field (i.e., $H_{\mathrm{D}}=-B\sum^N_{i=1}\hat{\sigma}^x_i$)}
to \YM{the problem Hamiltonian}
$H_{\mathrm{P}}$ over time as
% \YMdel{where $N$ is the number of the qubits.}
\begin{equation}
    H(t)= A_t H_{\mathrm{P}}+B_t H_{\mathrm{D}},
    \label{eq: total Hamiltonian}
\end{equation}
where $A_t$ and $B_t$ are the time-dependent coefficients.
% \begin{equation}
%     A_t = \dfrac{t}{T} \ (t:0 \rightarrow T) \nonumber
% \end{equation}
% \begin{equation}
%     B_t = 1 - \dfrac{t}{T} \ (t:0 \rightarrow T) \nonumber
% \end{equation}
The coefficients of the total Hamiltonian in QA are given by
\begin{gather}
    % H(t)= A_t H_{\mathrm{P}}+B_t H_{\mathrm{D}}, \\
    A_t = \dfrac{t}{T} \ (t:0 \rightarrow T), \nonumber \\
    B_t = 1 - \dfrac{t}{T} \ (t:0 \rightarrow T), \nonumber
    % \label{eq: total Hamiltonian}
\end{gather}
where $T$ is the annealing time (see Fig.~\ref{fig: conventional annealing schedule}). 
% \YMdel{We first have the initial state the ground state of $H_{\mathrm{D}}$.} 
If \YM{the} total Hamiltonian $H(t)$ is varied sufficiently slowly, 
% \YMdel{it is guaranteed by} 
the adiabatic theorem \YM{guarantees}
that we can obtain the ground state of $H_{\mathrm{P}}$.
}

\YS{
% \YMdel{Against ensuring the accuracy of QA, there are basically two impediments:}
In estimating the ground-state energy by QA,
there are two main problems:
environmental decoherence and non-adiabatic transitions~\cite{morita2008mathematical,albert1961quantum,albert1962quantum,roland2005noise,aaberg2005quantum,albash2015decoherence,childs2001robustness,sarandy2005adiabatic}. If we perform QA with a longer time schedule, we can avoid the effect of non-adiabatic transitions. 
% \YMdel{Actually, we cannot increase the time schedule unlimitedly since the quantum state is affected by the decoherence during the long time schedule.}
\YM{However, the longer time schedule makes quantum states more fragile against decoherence.}
This trade-off relation leads to the difficulty to solve practical problems with QA. 
}

\YS{There are many types of research to suppress the non-adiabatic transitions and decoherence during QA. 
% \YMdel{
% A method is proposed for the acceleration of QA by an inhomogeneous driver Hamiltonian for a specific case.}
\YM{Inhomogeneous driver Hamiltonian can be used to accelerate QA for a specific problem Hamiltonian~\cite{susa2018exponential,susa2018quantum}.}
Seki et al. show that the performance of QA for certain types of problem Hamiltonian can be improved with "non-stoquastic" Hamiltonians 
% \YMdel{having}
\YM{that have}
negative off-diagonal matrix elements~\cite{seki2012quantum,seki2015quantum}.
% \YMdel{A robust method against non-adiabatic transitions to directly estimate the energy gap between the ground and first energies is also proposed.}
\YM{It is known that the energy gap between the ground state and the first excited state can be estimated in a robust way against non-adiabatic transitions~\cite{matsuzaki2021direct,mori2022evaluate}.}
Moreover, there are 
% \YMdel{quite a few research}
several
works to suppress environmental noise.
% \YMdel{An idea for QA incorporating error correction can suppress decoherence. }
\YM{An error correction with ancillary qubits can be adopted to suppress decoherence during QA~\cite{pudenz2014error}.}
A scheme with a decoherence-free subspace for QA is also known~\cite{suzuki2020proposal}. Spin lock techniques are beneficial 
% \YMdel{when using}
\YM{to use}
long-lived qubits for QA~\cite{chen2011experimental,nakahara2013lectures,matsuzaki2020quantum}. In addition, there are several
% \YMdel{efficient studies for}
\YM{schemes to improve the performance of}
QA by using non-adiabatic transitions and quenching~\cite{crosson2014different,goto2020excited,hormozi2017nonstoquastic,muthukrishnan2016tunneling,brady2017necessary,somma2012quantum,das2008colloquium,karanikolas2020pulsed} and degenerating two-level systems~\cite{watabe2020enhancing}. Variational methods have also been applied to QA to suppress non-adiabatic transitions and decoherence~\cite{susa2021variational,matsuura2021variationally,passarelli2021transitionless,imoto2021quantum}.}

\section{Dual-state purification}
In this section, we review the EM method called dual-state purification~\cite{huo2022dual}. Firstly, we introduce VD/EES methods for NISQ devices in this paragraph which dual-state purification is based on.
In the VD/EES methods~\cite{koczor2021exponential,huggins2021virtual,czarnik2021qubit}, by using two copies of a noisy state $\rho$, we obtain a purified state $\rho_{\rm{VD/EES}}^{(2)}=\rho^2/\mathrm{Tr}[\rho^2]$.
\YM{Let us assume that the state is written by orthogonal basis
as $\rho = \sum_{n=0}^{2^N-1}p_n|\psi _n \rangle \langle \psi _n|$ where $|\psi _0\rangle =U|\vec{\bold{0}}\rangle $ denotes the ideal state, $U$ is a unitary operator without any noise, and $\ket*{\vec{\bold{0}}}$ denotes an initial state. In this case, the purified state $\rho_{\rm{VD/EES}}^{(2)}$ is closer to the ideal state than the original state for $p_0>1/2$.}
% \YMdel{This state is closer to the ideal final state $\ketbra*{\psi}=U\ketbra*{\vec{\bold{0}}}U^\dag$ where $U$ is a unitary operator without any noise, and $\ket*{\vec{\bold{0}}}$ denotes an initial state.}
The expectation value of an observable $O$ is estimated as $\ev{O}=\bra{\psi_0}O\ket{\psi_0}\simeq\mathrm{Tr}[O\rho^2]/\mathrm{Tr}[\rho^2]$.
\YM{However, in this scheme, the necessary number of qubits becomes twice larger than that of the original scheme without VD/EES methods.}

\YM{Let us review another EM scheme called dual-state purification that can be implemented by the same number of the qubits as the original scheme~\cite{huo2022dual}. To implement the dual-state purification,}
% \YMdel{To realize this estimation in dual-state purification, }
we consider a noisy map $\mathcal{F}(\bullet)=\sum_k F_k \bullet F_k^\dag$ where $\{F_i\}$ denote Kraus operators. When the noise amplitude is significantly small, this map is close to the ideal unitary operator $U$ (i.e., $\mathcal{F}(\bullet)\simeq U \bullet U^\dag$). 
% Dual-state purification requires 
% the inverse quantum operation of the noisy map $\mathcal{F}$. 
Let us consider a quantum process corresponding to the inverse quantum operation $U^{\dag}$ if there is no decoherence.
% \textcolor{red}{If there is no decoherence, we can implement the inverse quantum operation $U^{\dag}$.}
% \YMdel{Importantly, the inverse quantum operation should correspond to $U^\dag$ if there is no noise.}
% \YMdel{that can realize the inverse unitary operator $U^\dag$ if this mapping is an error-free process. }
We denote the quantum process
% \YMdel{This} 
with noise 
% \YMdel{is defined} 
as $\mathcal{G}(\bullet)=\sum_k G_k \bullet G_k^\dag$ where $\{G_i\}$ also denote Kraus operators. If the noise effect can be ignored, this noisy map $\mathcal{G}(\bullet)$ \YM{can be approximated as}
% \YMdel{approximates} 
$U^\dag \bullet U$. 

By using these operations $\mathcal{F}$ and $\mathcal{G}$, we obtain
\begin{eqnarray}
&&\bra*{\vec{\bold{0}}}\mathcal{G}\left(\mathcal{F}(\ketbra*{\vec{\bold{0}}})\right)\ket*{\vec{\bold{0}}} \nonumber \\
&=& \mathrm{Tr}[\bar{\mathcal{G}}(\ketbra*{\vec{\bold{0}}})\mathcal{F}(\ketbra*{\vec{\bold{0}}})]=\mathrm{Tr}[\bar{\rho}\rho],
\label{eq: dual-state purification}
\end{eqnarray}
where $\bar{\mathcal{G}}(\bullet)=\sum_k G_k^\dag \bullet G_k$ and 
% \YMdel{the state}
$\bar{\rho}=\bar{\mathcal{G}}(\ketbra*{\vec{\bold{0}}})$.
\YM{We call $\bar{\mathcal{G}}$  the dual map of  $\mathcal{G}$, while we call $\bar{\rho}$ the dual state.}
% \YMdel{are called the dual map of $\mathcal{G}$ and the dual state of $\rho$, respectively.}
We can obtain $\bar{\mathcal{G}}(\bullet)\simeq U \bullet U^\dag$ and $\mathrm{Tr}[\bar{\rho}\rho]\simeq \mathrm{Tr}[\rho^2]$ when the noise strength is sufficiently small.

\YM{Let us show how dual-state purification increases the population of the ideal state $\ket{\psi_0}$.}
% \YMdel{Let us check how this method improves the population of the target state $\ket{\psi}$ for a simple example.}
In this paragraph, we assume $\rho=(1-p)\ketbra*{\psi_0}+p\rho_e$ and $\bar{\rho}=(1-\bar{p})\ketbra*{\psi_0}+\bar{p}\bar{\rho}_e$ 
\YM{where $\rho_e$ ($\bar{\rho}_e$) denotes a normalized positive-semidefinite state
% after the map $\mathcal{G}$ $(\bar{\mathcal{G}})$ to the initial state 
and
$p$ ($\bar{p}$) denotes an error probability to satisfy $1-p>p$  ($1-\bar{p}>\bar{p}$).}
% \YMdel{where $p$ and $\bar{p}$ denote error probabilities and satisfy $1-p>p$ and $1-\bar{p}>\bar{p}$, and $\rho_e$ and $\bar{\rho}_e$ denote normalized positive-semidefinite states.}
\YM{
% When the state is affected by random Pauli noise, 
Also we assume incoherent errors where the state 
$\rho_e$ ($\bar{\rho}_e$) satisfies $\ketbra*{\psi_0}\rho_e=0$ ($\ketbra*{\psi_0}\bar{\rho}_e=0$)~\cite{koczor2021exponential,huggins2021virtual}. 
}
% \YMdel{these states $\rho_e$ and $\bar{\rho}_e$ satisfy $\ketbra*{\psi}\rho_e=\ketbra*{\psi}\bar{\rho}_e=0$}
% \YMdel{when the states are derived from incoherent errors. }
The virtually purified state without normalization is defined as follows:
\begin{equation}
    \dfrac{\rho\bar{\rho} + \bar{\rho}\rho}{2}
    = (1-p)(1-\bar{p}) \ketbra*{\psi_0} + p\bar{p}\ \mathrm{Tr}[\rho_e\bar{\rho}_e] \dfrac{\rho_e\bar{\rho}_e + \bar{\rho}_e\rho_e}{2\mathrm{Tr}[\rho_e\bar{\rho}_e]},
    \label{eq:virtually-obtained state}
\end{equation}
where the state $\rho\bar{\rho} + \bar{\rho}\rho$ is Hermitian. Note that this EM method decreases 
\YS{the ratio between the error state and the ideal state from $p/(1-p)$ 
and $\bar{p}/(1-\bar{p})$ 
to $p\bar{p} \ \mathrm{Tr}[\rho_e\bar{\rho}_e]/(1-p)(1-\bar{p}) \leq p\bar{p}/(1-p)(1-\bar{p})$. 
By considering the state $\rho\bar{\rho} + \bar{\rho}\rho$, this EM method can effectively improve the population of the ideal state without the information of the noise model unlike previous schemes such as the quasi-probability method.}
% \YMdel{the population of the error state from $p$ and $\bar{p}$ to $p\bar{p}$.}

\begin{figure}[htbp]
  \begin{minipage}[]{1\linewidth}
  \centering
  \subcaption{\raggedright}
  \includegraphics[clip,width=0.8\textwidth]{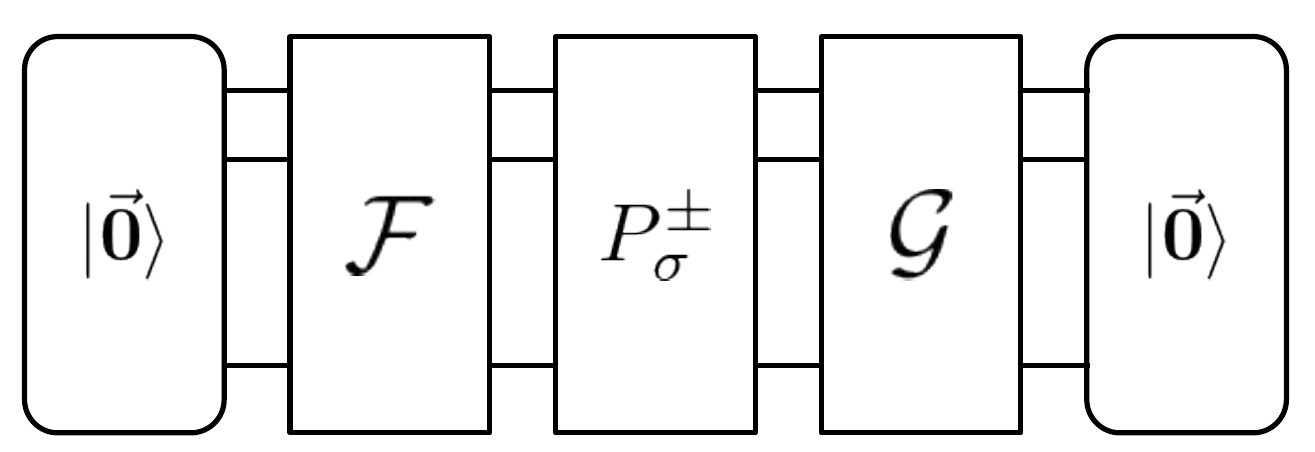}
  \label{fig: error mitigation circuit for numerator}
  \end{minipage}
  \begin{minipage}[]{1\linewidth}
  \centering
  \subcaption{\raggedright}
  \includegraphics[clip,width=0.6\textwidth]{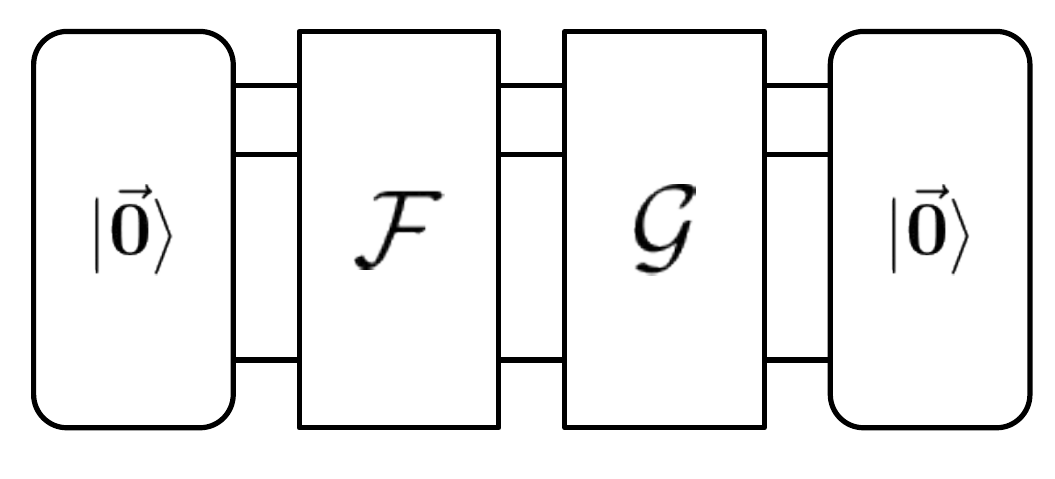}
  \label{fig: error mitigation circuit for denominator}
  \end{minipage}
  \caption{The circuits for implementing dual-state purification. We calculate the numerator (the denominator) in Eq.~\eqref{eq: expectation value} with the circuit in Fig.~\ref{fig: error mitigation circuit for numerator} (Fig.~\ref{fig: error mitigation circuit for denominator}). First, we perform a noisy map $\mathcal{F}$ in Eq.~\eqref{eq: dual-state purification} to an initial state $\ket*{\vec{\bold{0}}}$. 
  Second, when we
  calculate the numerator in Eq.~\eqref{eq: expectation value}, we perform the projective measurement $P_{\sigma}^{\pm}$ of a Pauli product $\sigma$ as shown in Fig.~\ref{fig: error mitigation circuit for numerator}. For obtaining the denominator in Eq.~\eqref{eq: expectation value} by the circuit in Fig.~\ref{fig: error mitigation circuit for denominator}, the projective measurement is not implemented. Third,
  we perform a noisy inverse map $\mathcal{G}$ in Eq.~\eqref{eq: dual-state purification}. Finally, we measure the state on the basis of the initial state $\ket*{\vec{\bold{0}}}$.}
  \label{fig: error mitigation circuit}
\end{figure}

We obtain the expectation value of an observable $O$ as $\ev{O}=\Sigma a_i\ev{\sigma_i}$ by decomposing $O$ as the summation of the expectation values of Pauli products $\{\sigma_i\}$ where $a_i$ is a coefficient. 
Dual-state purification gives the expectation value of  a Pauli product $\sigma$ as
\begin{equation}
    \ev{\sigma} \simeq \left. \mathrm{Tr} \left [ \sigma \dfrac{\rho\bar{\rho} + \bar{\rho}\rho}{2}\right] \middle/ \mathrm{Tr} \left [\dfrac{\rho\bar{\rho} + \bar{\rho}\rho}{2}\right] \right.
    \label{eq: expectation value}
\end{equation}
where $\mathrm{Tr}\left [\dfrac{\rho\bar{\rho} + \bar{\rho}\rho}{2}\right]$ denotes a normalization factor.
% \YMdel{The circuit in Fig.~\ref{fig:error mitigation} provides this expectation. }
\YM{To obtain this expectation value, we can implement quantum circuits described in Fig.~\ref{fig: error mitigation circuit}.}
We perform the circuit in Fig.~\ref{fig: error mitigation circuit for numerator} (Fig.~\ref{fig: error mitigation circuit for denominator}) to obtain the numerator (the denominator) in Eq.~\eqref{eq: expectation value}.
First, prepare an initial state $\ket*{\vec{\bold{0}}}$, and let this state evolve by the noisy quantum circuit described by the channel  $\mathcal{F}$.
% \YMdel{the quantum state starts from the initial state $\ket*{\vec{\bold{0}}}$, and then the noisy operation $\mathcal{F}$ is implemented. }
Second, 
perform a projective measurement $P_{\sigma}^{\pm}$ of a Pauli product $\sigma$
for the numerator as shown in Fig.~\ref{fig: error mitigation circuit for numerator}
% \YMdel{at the gray box Fig.~\ref{fig:error mitigation}}
where the measurement operator corresponding
to the outcome $\pm1$ is given by
$P_{\sigma}^{\pm}=(I \pm \sigma)/2$.
% \YMdel{if we choose the identity operator (the projective measurement $P_O^{\pm}$ of an observable $O$) at the gray box, we can compute the denominator (the numerator) in Eq.~\ref{eq: expectation value} where $P_{O}^{\pm}=(I \pm O)/2$.}
When we implement the circuit in Fig.~\ref{fig: error mitigation circuit for denominator} to obtain the denominator in Eq.~\eqref{eq: expectation value}, we do not perform the projective measurement.
\YM{Third, let the state evolve by the noisy quantum process described by the channel $\mathcal{G}$.}
% \YMdel{Then, the inverse map operates the quantum state.}
\YM{Finally, we investigate how much population remains in $\ket*{\vec{\bold{0}}}$
by a measurement in the basis including $\ket*{\vec{\bold{0}}}$.}
% \YMdel{Finally, the quantum state is measured to investigate how population the initial state $\ket*{\vec{\bold{0}}}$ occupies in the final state.}
% \YMdel{The denominator in Eq.~\ref{eq: expectation value} is given by the circuit in Fig.~\ref{fig:error mitigation} without the projective measurement, i.e.,}
\YM{If we choose the circuit in Fig.~\ref{fig: error mitigation circuit for denominator}, we can measure the denominator in Eq.~\eqref{eq: expectation value} as follows:}
\begin{equation}
    P_{\vec{\bold{0}}}=\bra*{\vec{\bold{0}}}\mathcal{G}\left(\mathcal{F}(\ketbra*{\vec{\bold{0}}})\right)\ket*{\vec{\bold{0}}} =\mathrm{Tr}[\bar{\rho}\rho],
\end{equation}
where $P_{\vec{\bold{0}}}$ denotes the population of $\ket*{\vec{\bold{0}}}$ in the final state obtained by the measurement.
\YM{On the other hand, if we choose the circuit in Fig.~\ref{fig: error mitigation circuit for numerator}, we can measure the numerator in Eq.~\eqref{eq: expectation value} as}
% \YMdel{By using the circuit in Fig.~\ref{fig:error mitigation} with the projective measurement, we can calculate the numerator in Eq.~\ref{eq: expectation value} as} 
$\tilde{P}_{\vec{\bold{0}}}^{+}-\tilde{P}_{\vec{\bold{0}}}^{-}$ where
\begin{equation}
    \tilde{P}_{\vec{\bold{0}}}^{\pm}=\bra*{\vec{\bold{0}}}\mathcal{G}\left(P_{\sigma}^{\pm}\mathcal{F}(\ketbra*{\vec{\bold{0}}})P_{\sigma}^{\pm}\right)\ket*{\vec{\bold{0}}} =\mathrm{Tr}[P_{\sigma}^{\pm}\rho P_{\sigma}^{\pm}\bar{\rho}],
    \label{eq: expected value}
\end{equation}
$\tilde{P}_{\vec{\bold{0}}}^{\pm}$ also denote the population of $\ket*{\vec{\bold{0}}}$.
To sum up, the expectation value is rewritten by
\begin{equation}
    \ev{\sigma} = (\tilde{P}_{\vec{\bold{0}}}^{+}-\tilde{P}_{\vec{\bold{0}}}^{-})/P_{\vec{\bold{0}}}.\label{eq: expectation of dual state purification}
\end{equation}

\YS{In EM, we also need to suppress the sampling noise which leads to the residual error. We can decrease the variance of the expectation value in Eq.~\eqref{eq: expectation of dual state purification} by increasing
\YM{the number of measurements.}
% \YMdel{circuit shots.} 
\YM{Due to the denominator,
the variance of Eq.~\eqref{eq: expectation of dual state purification} is larger than that of Eq.~\eqref{eq: expected value}. Especially, when
$\mathrm{Tr}[\bar{\rho}\rho]$ approaches to $0$, the variance diverges to an infinity. So, in order to implement our scheme within a finite time, we need a finite value of $\mathrm{Tr}[\bar{\rho}\rho]$.
}
% \YMdel{
% This variance can be amplified by the denominator $P_{\vec{\bold{0}}}$, which means the expected value is not calculable when the denominator is close to zero. Thus, this method requires generating states to have finite fidelity, which makes the denominator finite.}
}

\section{Error-mitigated quantum annealing}
In this section, we describe our QA scheme to obtain the ground-state energy of a problem Hamiltonian by mitigating environmental noise effects. To adapt the EM method mentioned in the previous section to QA, we design QA scheduling to construct the operations $\mathcal{F}$ and $\mathcal{G}$. 
% \YMdel{At first,}
\YM{Firstly,}
we explain a naive schedule 
% \YMdel{derived from} 
\YM{inspired by} the reverse quantum annealing (RQA). 
\YS{In an ideal situation, RQA can let the quantum state evolve from the initial state to the ground state of the problem Hamiltonian and then evolve to the final state close to the initial state. 
Thus, RQA seems to provide the inverse map. However, this schedule cannot purify the noisy state if non-adiabatic transitions can occur as we \YM{will explain} below.}

\YM{Secondly, we introduce another scheduling to overcome such a problem. We call this approach error-mitigated quantum annealing (EMQA).
Actually, in this scheme, the inverse map can be realized for any initial states
as long as there is no decoherence. Unlike the naive schedule with RQA, EMQA provides the inverse map even when non-adiabatic transitions occur. 
We show that this approach provides a more accurate estimation of the energy than the conventional approach.}
% \YMdel{
% Though the reverse quantum annealing seems to provide the inverse map, this schedule cannot purify the noisy state. We show that this schedule gives a lower energy than the ground-state energy in the next section. 
% Moreover, we theoretically explain why we cannot obtain the correct ground-state energy from this schedule in Appendix(??).
% Then, we introduce another schedule to realize the purified state. In this schedule, we can compute the more accurate ground-state energy than in the conventional one and the naive one.}

\begin{figure}[htbp]
  \begin{minipage}[]{1\linewidth}
  \centering
  \subcaption{\raggedright}
  \includegraphics[clip,width=0.7\textwidth]{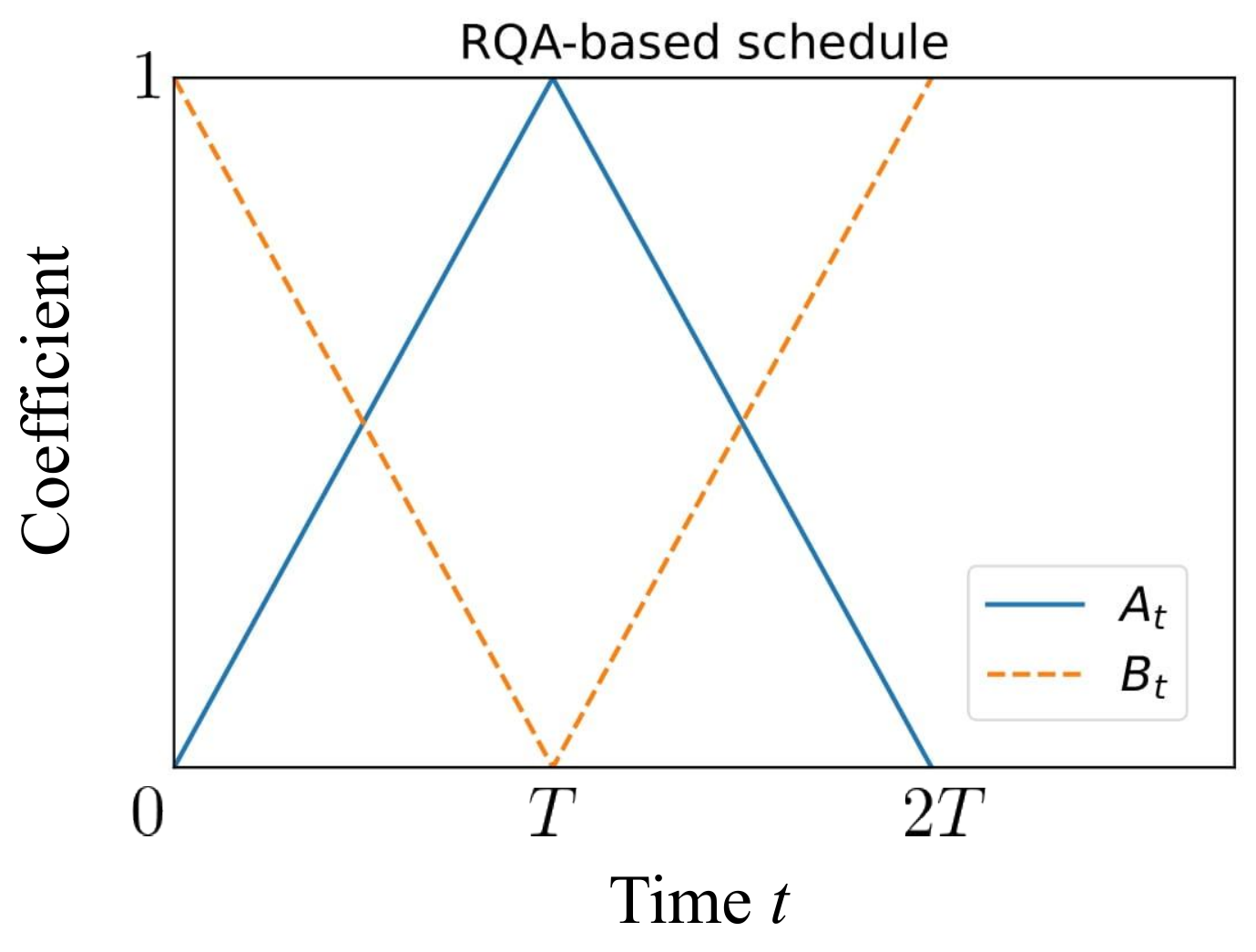}
  \label{fig: RQA-based schedule}
  \end{minipage}
  \begin{minipage}[]{1\linewidth}
  \centering
  \subcaption{\raggedright}
  \includegraphics[clip,width=0.8\textwidth]{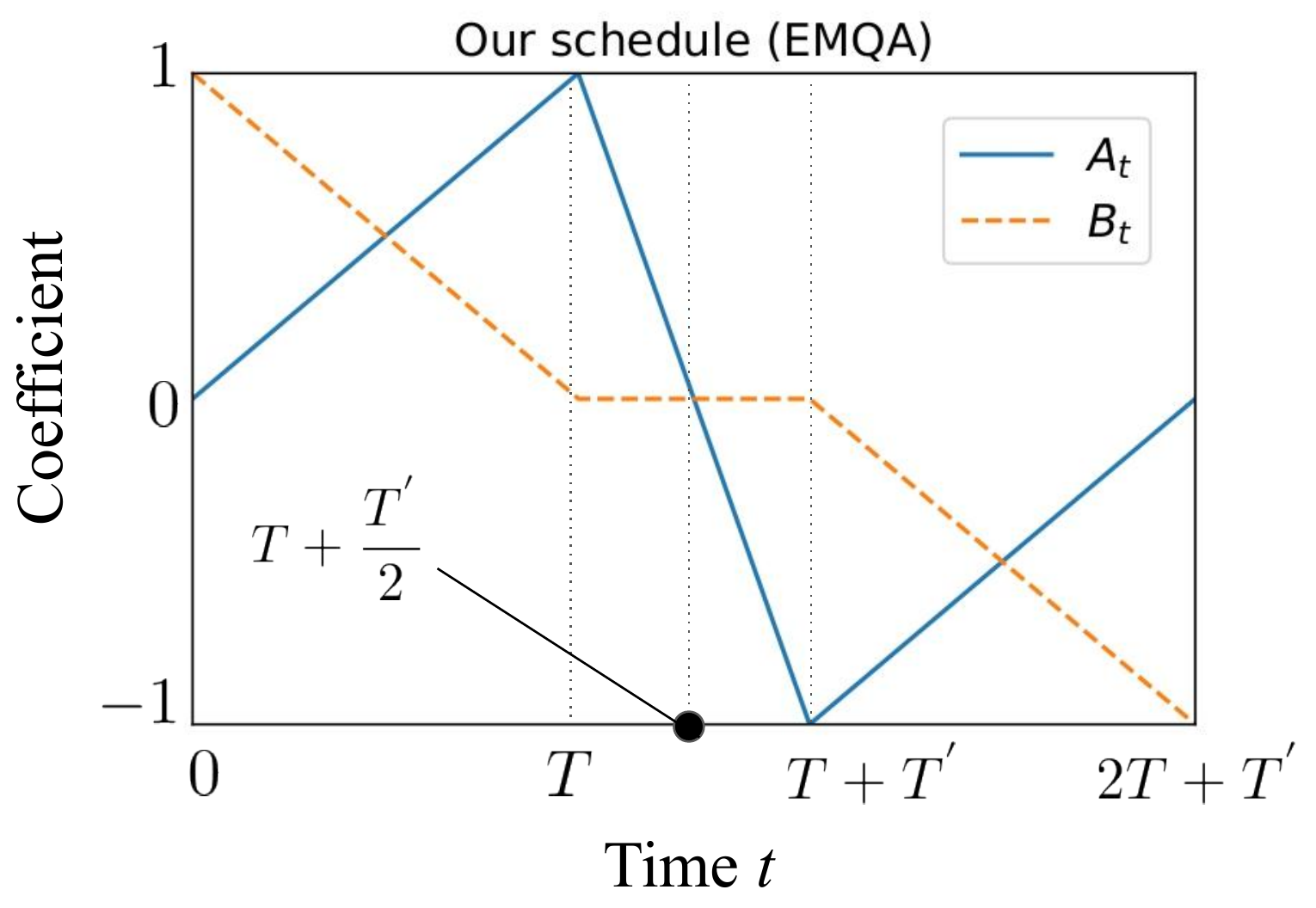}
  \label{fig: schedule of error-mitigated QA}
  \end{minipage}
  \caption{Two schedules of QA to implement the EM method called dual-state purification. (a)
  A schedule 
  % \YMdel{in Fig.~\ref{fig: RQA-based schedule}} \YMdel{derives from} 
  \YM{inspired by} 
  RQA. The dynamics from $t=0$ to $t=T$ ($t=T$ to $t=2T$) corresponds to the noisy map $\mathcal{F}$ ($\mathcal{G}$) in Eq.~\eqref{eq: dual-state purification}. At $t=T$, we perform a projective measurement to calculate the numerator in Eq.~\eqref{eq: expectation value}. In Sec.~\ref{sec: results}, we show that we cannot 
  % \YMdel{execute accurate estimation for} 
  \YM{accurately estimate}
  the ground-state energy of $H_{\mathrm{P}}$ with this schedule.
  % \YMdel{The}
  % \YMdel{In Fig.~\ref{fig: schedule of error-mitigated QA},}
  (b)
  \YM{An improved}
  schedule  
  % \YMdel{with} 
\YM{tailored for} the EM method.
% \YMdel{\YM{This} gives a more precise estimation than \YM{both}
% the RQA-based and conventional schedules in decoherence. 
% }
The dynamics from $t=0$ to $t=T+T^{'}/2$ corresponds to the noisy map $\mathcal{F}$, and this schedule from $t=T+T^{'}/2$ to $t=2T+T^{'}$ 
\YM{provides a dynamics corresponding to}
the inverse map $\mathcal{G}$ in Eq.~\eqref{eq: dual-state purification}. For calculating the numerator in Eq.~\eqref{eq: expectation value}, a projective measurement is needed at $t=T+T^{'}/2$.}
\end{figure}

% \YMdel{In the naive schedule,}
\YM{Let us show the RQA-based schedule.}
The 
% \YMdel{total-Hamiltonian} 
coefficients in Eq.~\eqref{eq: total Hamiltonian} are written by
\begin{equation}
A_t=\left\{
\begin{array}{ll}
\dfrac{t}{T} & (t:0\rightarrow T) \\
-\dfrac{t}{T} + 2 & (t:T \rightarrow 2T),
\end{array}
\right.
\end{equation}
\begin{equation}
B_t=\left\{
\begin{array}{ll}
-\dfrac{t}{T} + 1 & (t:0\rightarrow T) \\
\dfrac{t}{T} - 1 & (t:T \rightarrow 2T).
\end{array}
\right.
\end{equation}
This annealing schedule \YM{is described in} Fig.~\ref{fig: RQA-based schedule}. 
% \YMdel{To obtain the expected value, }
\YM{To obtain the numerator of the Eq.~\eqref{eq: expectation of dual state purification} with dual-state purification,}
we \YM{need to}
implement a projective measurement \YM{$P_{\sigma}^{\pm}$}
% \YMdel{of an observable $O$} 
at $t=T$. The dynamics from $t=0$ to $t=T$ corresponds to the map $\mathcal{F}$ while the dynamics from $t=T$ to $t=2T$ corresponds to the inverse map $\mathcal{G}$.
\YM{Here, let us \YS{consider the specific case in which} the following three conditions are satisfied. First, the adiabatic condition is satisfied. Second, the initial state is the ground state of the driver Hamiltonian. Third, there is no decoherence. If these conditions are satisfied,
%since this schedule is based on RQA, we can realize
the final state returns back to the initial state. This seems to suggest that the RQA-based schedule provides 
the inverse map, and dual-state purification by using this schedule
may work in practical circumstances. 
%if these conditions are satisfied.
%without decoherence 
%when the initial state is the ground state of the Hamiltonian.
However, we show that, if there are non-adiabatic transitions, the RQA-based schedule does not provide the proper inverse map. Actually, in this schedule, the virtually-obtained state described by Eq.~\eqref{eq:virtually-obtained state} could be unphysical because the energy of the state could be lower than the ground-state energy of the problem Hamiltonian. We will explain the origin of this unphysicality with both analytical and numerical methods.}

\YS{On the other hand, the more promising scheme, EMQA, is described by 
\begin{equation}
A_t=\left\{
\begin{array}{ll}
\dfrac{t}{T} & (t:0\rightarrow T) \\
-\dfrac{2t}{T^{'}} + \dfrac{2T}{T^{'}} + 1 & (t:T \rightarrow T + T^{'}) \\
\dfrac{t}{T} - \dfrac{T^{'}}{T} - 2 & (t:T + T^{'}\rightarrow 2T + T^{'}),
\end{array}
\right.
\label{eq: coefficient A of EMQA}
\end{equation}
\begin{equation}
B_t=\left\{
\begin{array}{ll}
-\dfrac{t}{T} + 1 & (t:0\rightarrow T) \\
0 & (t:T \rightarrow T + T^{'}) \\
-\dfrac{t}{T} + \dfrac{T^{'}}{T} + 1 & (t:T + T^{'}\rightarrow 2T + T^{'}).
\end{array}
\right.
\label{eq: coefficient B of EMQA}
\end{equation}
This schedule draws Fig.~\ref{fig: schedule of error-mitigated QA}.
From $t=T$ to $t=T+T^{'}$, if there is 
% \YMdel{the noise effect}
\YM{decoherence}
from the environment, we should \YM{instantaneously change}
% \YMdel{switch} 
the sign of the problem Hamiltonian.
% \YMdel{in a moment. However, any quantum devices basically do not allow us to change the sign immediately. Therefore, we should set $T^{'}$ as short as the devices permit.}
\YM{However, such an immediate change in the sign of the Hamiltonian
is difficult to implement for the QA devices. So we keep the time as short as possible within the bandwidth of the devices.
}
% By performing the projective measurement of an observable $O$ at $t=T+\dfrac{T^{'}}{2}$, we can obtain the expected value.
\YM{Again, to obtain the numerator of the Eq.~\eqref{eq: expectation of dual state purification}, 
we need to perform a projective measurement \YM{$P_{\sigma}^{\pm}$} at $t=T+T^{'}/2$.}
Thus, the dynamics from $t=0$ to $t=T+T^{'}/2$ represents the map $\mathcal{F}$, while the dynamics from $t=T+T^{'}/2$ to $t=2T+T^{'}$ 
% \YMdel{means}
\YM{corresponds to}
the inverse map $\mathcal{G}$. We show in Appendix A that the time evolution from $t=T+T^{'}/2$ to $t=2T+T^{'}$ is 
% \YMdel{truly equal}
\YM{actually equivalent}
to the
conjugate transpose of the dynamics from $t=0$ to $t=T+T^{'}/2$ when there is no decoherence even if the adiabatic condition is not satisfied.
}

\YM{To take into account decoherence, we adopt the Gorini–Kossakowski–Sudarshan–Lindblad (GKSL) master equation to describe the dynamics of the system}
\YS{
% \YMdel{During the time evolution, we assume the dynamics of the density matrix $\rho$ is described by the GKSL equation as follows:}
\begin{equation}
    \dfrac{d\rho}{dt} = -i[H(t), \rho]+\sum_{n}\dfrac{\lambda_{n}}{2}\left(2\hat{L}_n \rho \hat{L}^\dag_n - \left \{ \hat{L}_n^\dag \hat{L}_n, \rho \right \}
    \right),
\end{equation}
where $[\bullet,\bullet]$ denotes the commutator, $\lambda_{n}$ 
% \YMdel{is coefficients} 
\YM{denotes a decay rate}, $\hat{L}_n$ denotes the Lindblad operator, and $\{\bullet,\bullet\}$ denotes the anticommutator.}

% Throughout this paper, we assume the quantum device which can implement continuous operations, single-qubit rotations, and projective measurements but cannot perform two-qubit gates. This assumption has been discussed as digital-analog quantum computation (DAQC). This approach has been proposed as a hybrid architecture to realize the flexibility of NISQ computation on robust analog quantum simulators.
Throughout this paper, \YM{we assume that we can realize both the dynamics induced by the Hamiltonian and arbitrary single-qubit operations.
Due to the recent development of quantum technologies, such a device is feasible, which we will discuss later.}
% \YMdel{we assume the quantum device which can implement QA and single-qubit operations but cannot perform two-qubit gates.
% Owing to the recent quantum technology, such a quantum device can be realized in the near future, which we will discuss later.}
However, 
\YM{it is unclear whether we can implement two-qubit gates with high fidelity, and so we assume that we cannot perform two-qubit gates on the devices.}
% \YMdel{it is not obvious whether the quantum device can perform two-qubit gates with high fidelity or not.}
This assumption is similar to 
\YM{that of digital-analog quantum computation (DAQC)~\cite{parra2020digital}.}
% \YMdel{the counterpart discussed in the area of digital-analog quantum computation (DAQC). }
This approach has been proposed as a hybrid architecture to realize the flexibility of NISQ computation on robust analog quantum simulators.

It is worth mentioning that dual-state purification is 
% \YMdel{experimentally more feasible}
\YM{more suitable}
for
\YM{the device considered by us}
% \YMdel{QA} 
than virtual distillation. Although virtual distillation is also an efficient method to suppress stochastic errors, virtual distillation requires two-qubit gates to entangle $M$ copies of a noisy state $\rho$. 
% \YMdel{We cannot basically implement two-qubit gates with QA devices.}
\YM{Thus, it is not straightforward to implement virtual distillation under our assumption.}
On the other hand, in dual-state purification, we do not need to prepare $M$ copies of a noisy state $\rho$ or \YM{implement} two-qubit gates as we mentioned in the previous section. This means that EMQA is more practical to mitigate noise effects during QA.

\section{Results}
\label{sec: results}

\begin{figure*}
    \begin{tabular}{cc}
      \begin{minipage}[t]{0.5\hsize}
        \centering
        \includegraphics[clip,width=0.8\textwidth]{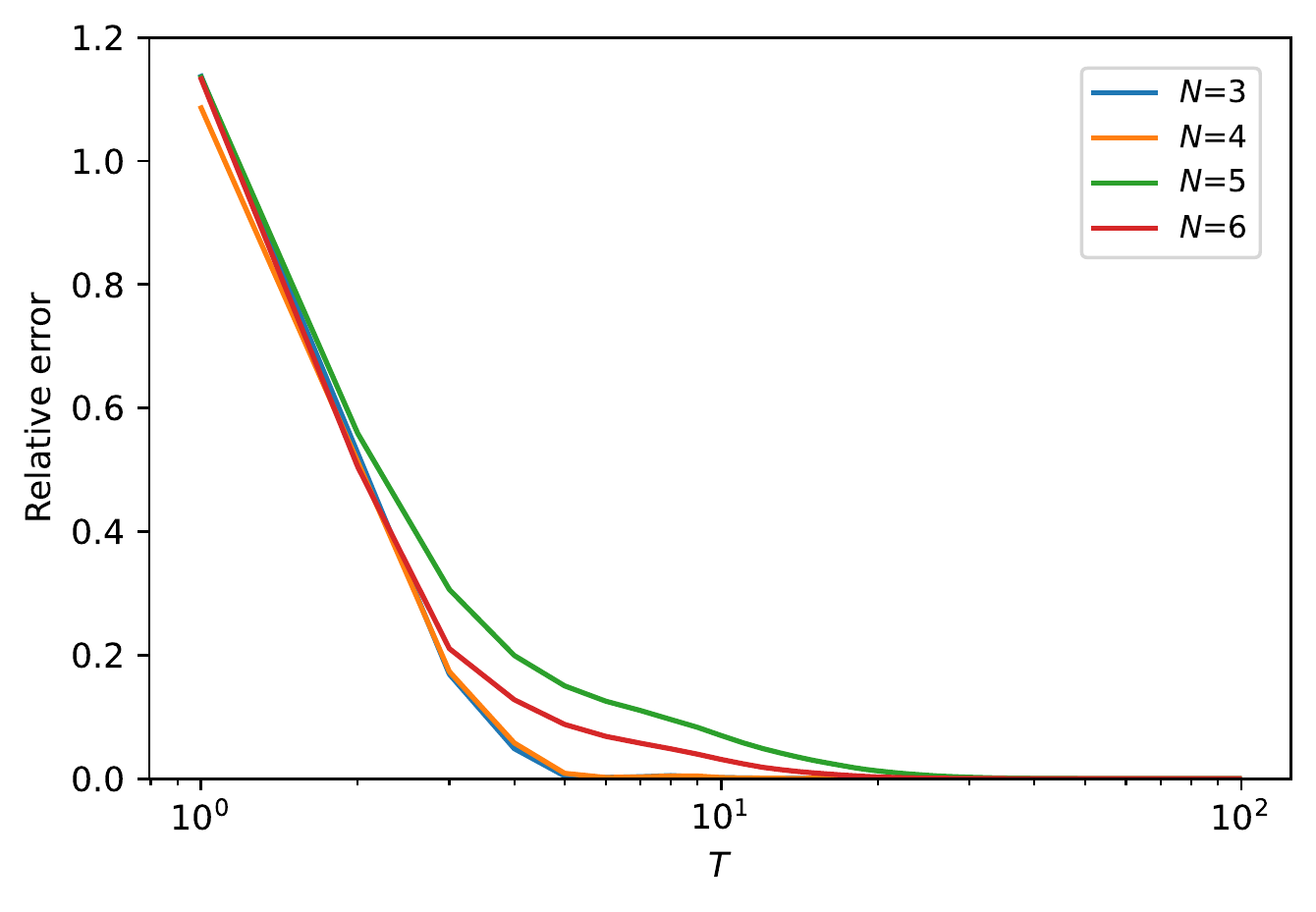}
        \subcaption{EMQA schedule $\lambda=0$}
        \label{fig: result of EMQA with lambda=0}
      \end{minipage} &
      \begin{minipage}[t]{0.5\hsize}
        \centering
        \includegraphics[clip,width=0.8\textwidth]{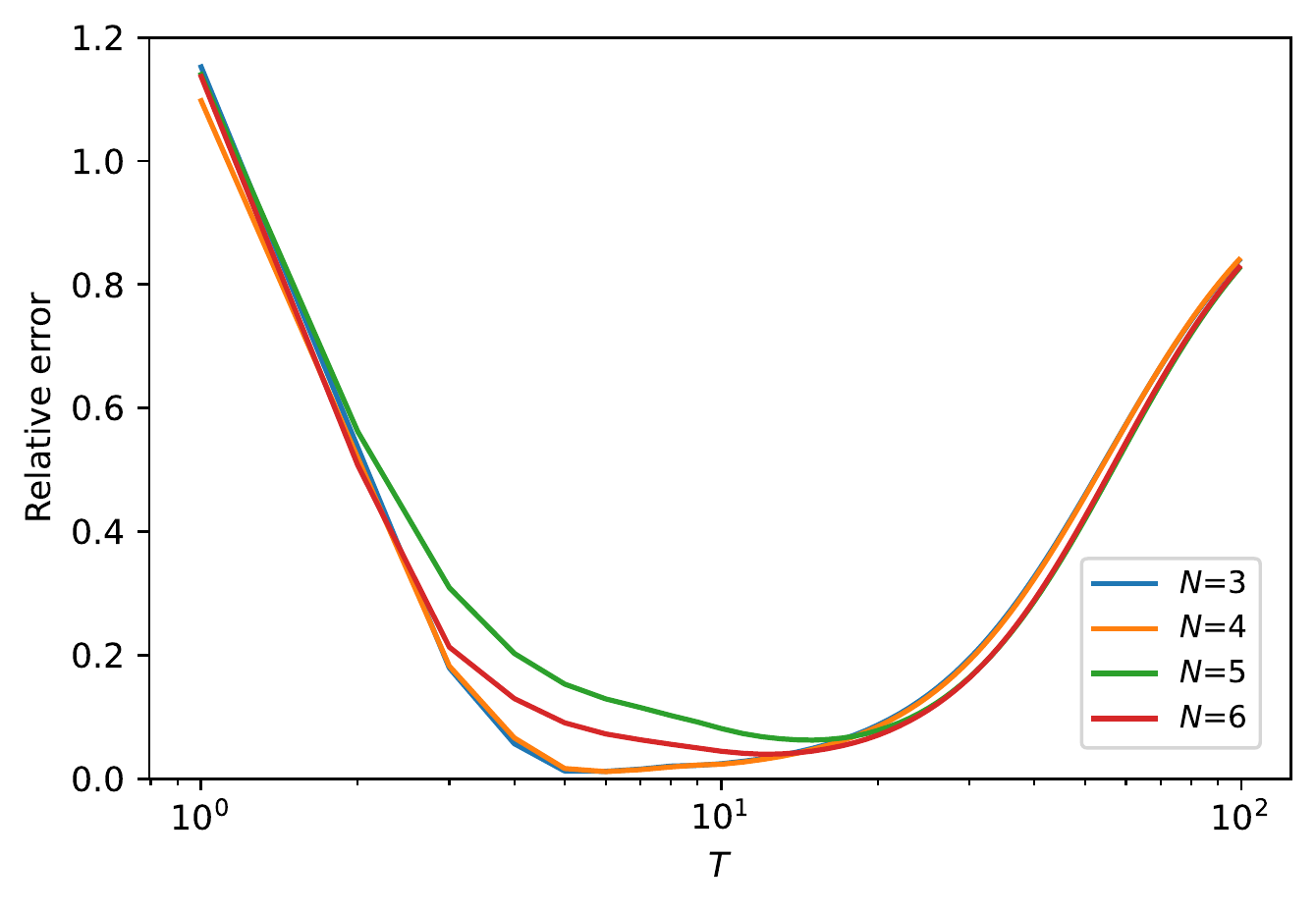}
        \subcaption{EMQA schedule $\lambda=0.004$}
        \label{fig: result of EMQA with lambda=0.004}
      \end{minipage} \\
   
      \begin{minipage}[t]{0.5\hsize}
        \centering
        \includegraphics[clip,width=0.8\textwidth]{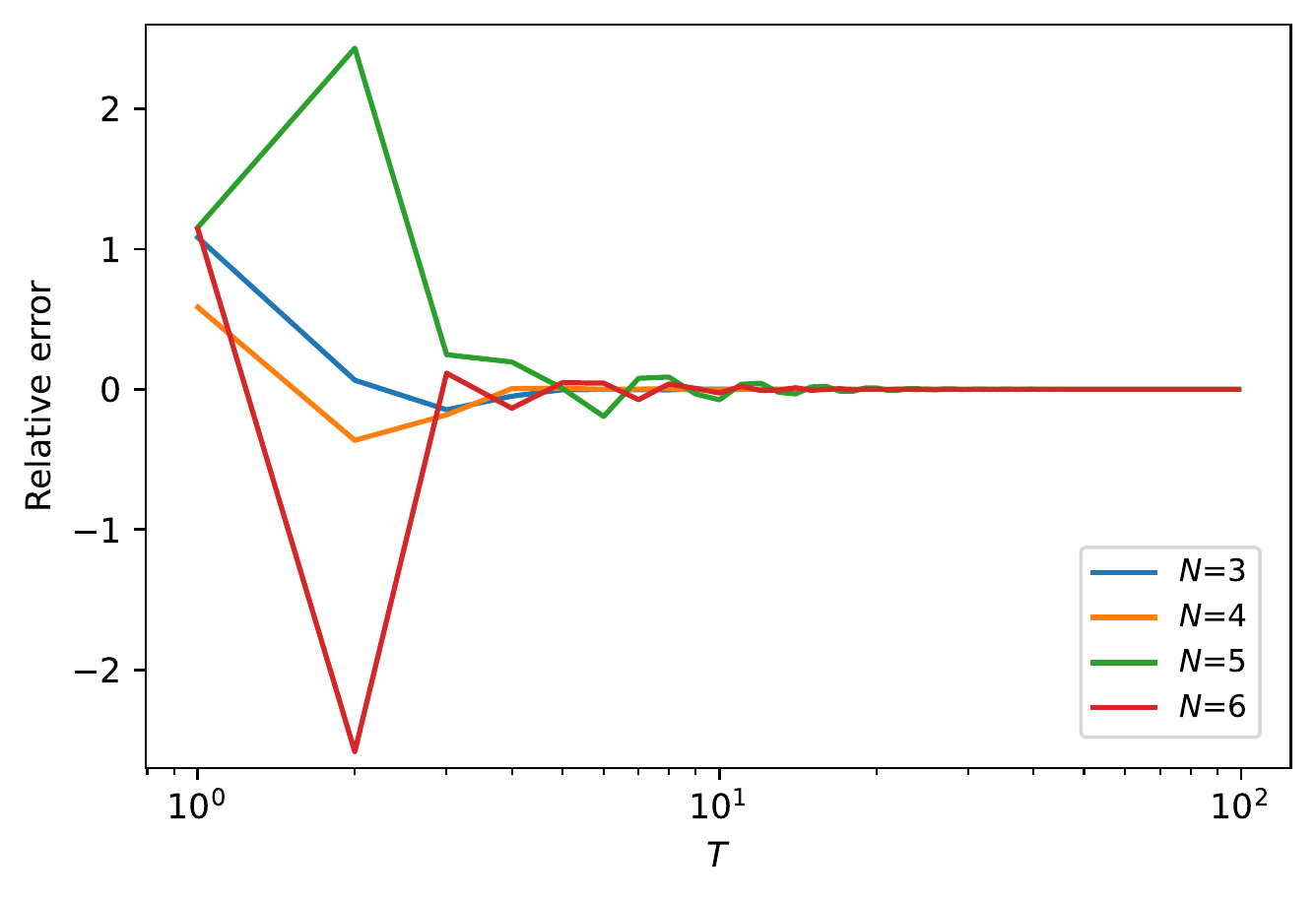}
        \subcaption{RQA-based schedule $\lambda=0$}
        \label{fig: result of RQA-based with lambda=0}
      \end{minipage} &
      \begin{minipage}[t]{0.5\hsize}
        \centering
        \includegraphics[clip,width=0.8\textwidth]{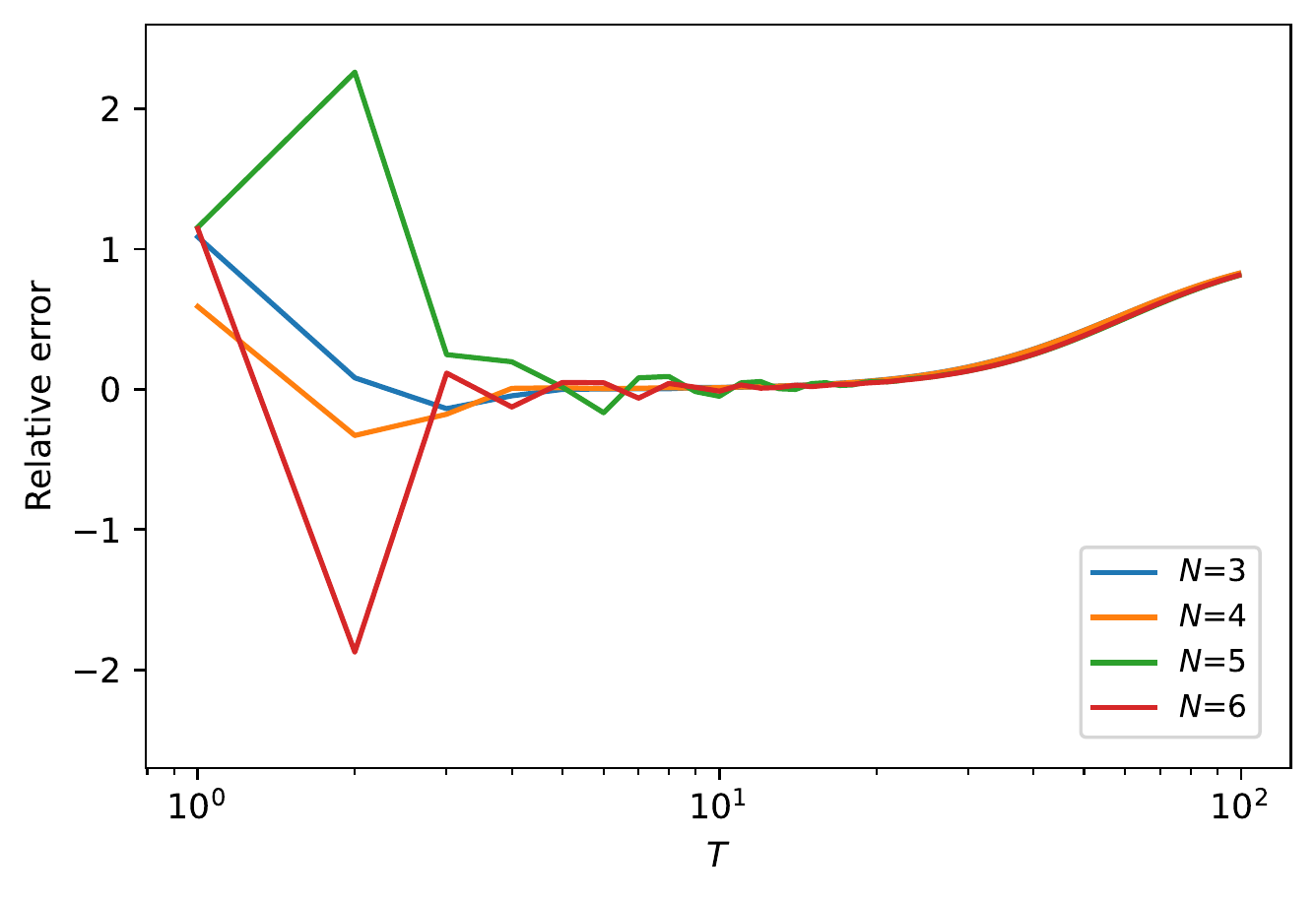}
        \subcaption{RQA-based schedule $\lambda=0.004$}
        \label{fig: result of RQA-based with lambda=0.004}
      \end{minipage} \\
      
      \begin{minipage}[t]{0.5\hsize}
        \centering
        \includegraphics[clip,width=0.8\textwidth]{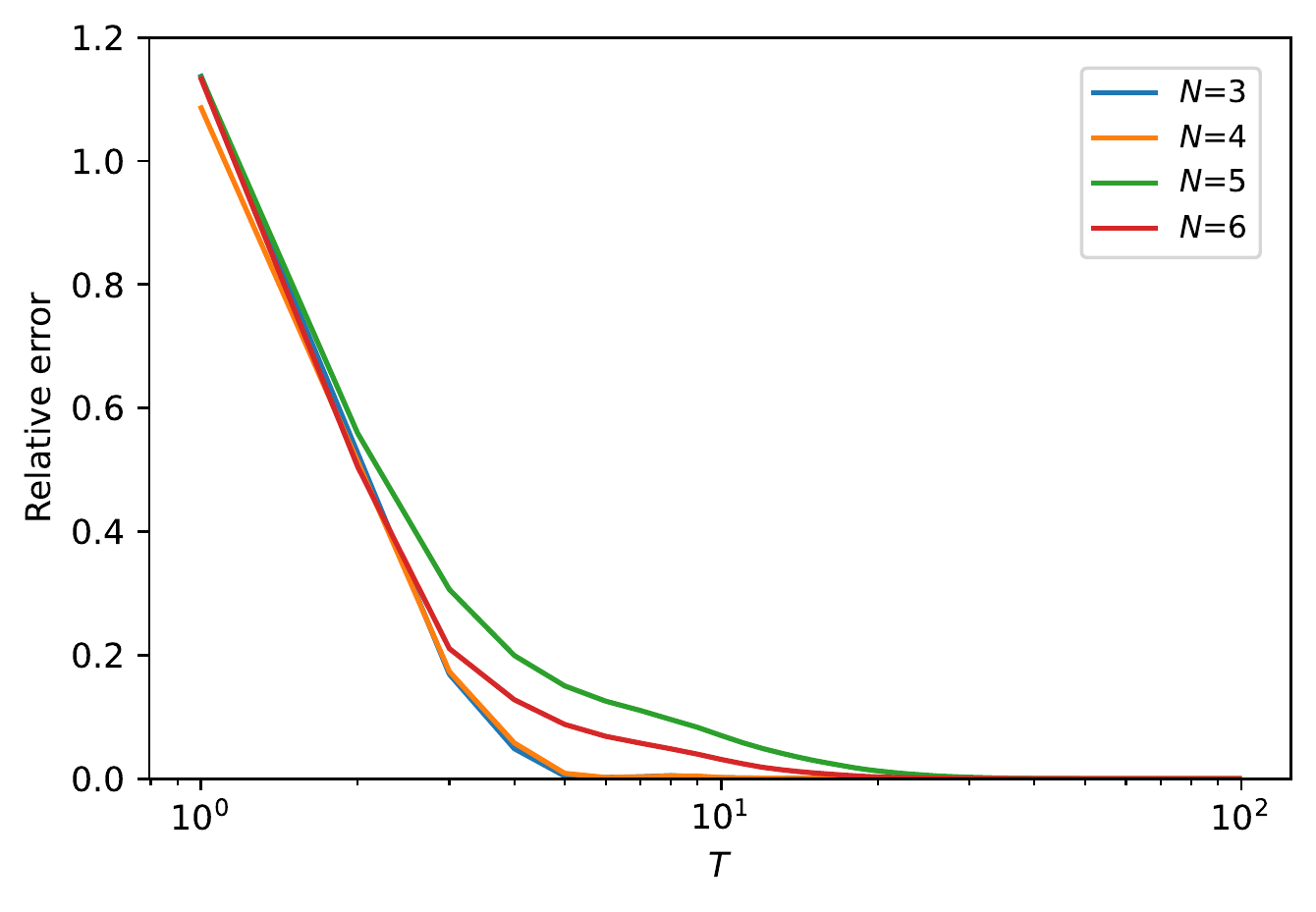}
        \subcaption{Conventional schedule $\lambda=0$}
        \label{fig: result of conventional QA with lambda=0}
      \end{minipage} &
      \begin{minipage}[t]{0.5\hsize}
        \centering
        \includegraphics[clip,width=0.8\textwidth]{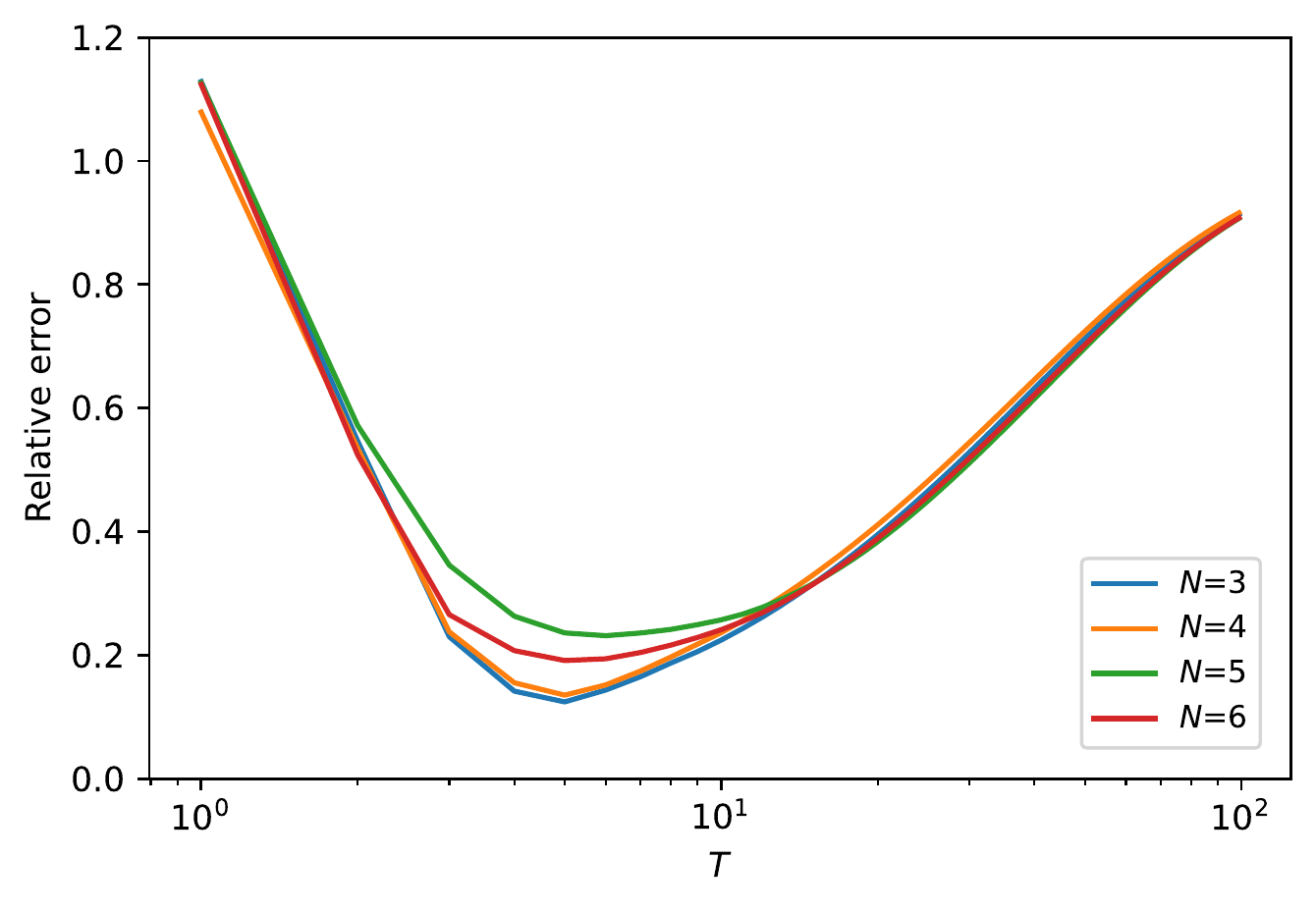}
        \subcaption{Conventional schedule $\lambda=0.004$}
        \label{fig: result of conventional QA with lambda=0.004}
      \end{minipage} 
    \end{tabular}
    \caption{
    % \YMdel{The relative error between}
    A relative error of the ground-state energy of the problem Hamiltonian $H_{\mathrm{P}}$. We define the relative error as
     $(\ev{H_{\mathrm{P}}}-E_{\mathrm{g}}) / \abs{E_{\mathrm{g}}}$
     where $\ev{H_{\mathrm{P}}}$ denotes the expectation value with QA and $E_{\mathrm{g}}$ denotes the actual value of the ground-state energy.
    We use
    (a) the EMQA, (c) RQA-based, and (e) conventional schedules with $\lambda=0$, while we adopt (b) the EMQA, (d) RQA-based, and (f) conventional schedules with $\lambda / J=0.004$. The annealing time $T$ is normalized by $J$.
    % \YMdel{ the expected value obtained by each method and the ground-state energy of the problem Hamiltonian $H_{\mathrm{P}}$ \YMdel{at each} \YM{against} annealing time $T$. \YMdel{on a log scale by changing the number of qubits $N$.} Fig.~\ref{fig: result of EMQA with lambda=0}, \ref{fig: result of RQA-based with lambda=0}, and \ref{fig: result of conventional QA with lambda=0} are the plots with the EMQA, RQA-based, and conventional schedules at $\lambda=0$, respectively. Fig.~\ref{fig: result of EMQA with lambda=0.004}, \ref{fig: result of RQA-based with lambda=0.004}, and \ref{fig: result of conventional QA with lambda=0.004} describe the results with the EMQA, RQA-based, and conventional schedules at $\lambda=0.004$, respectively.}
    We 
    % \YMdel{check} 
    \YM{show that} the EMQA schedule 
    % \YMdel{in Fig.~\ref{fig: result of EMQA with lambda=0.004}} \YMdel{leads to}
    \YM{provides the minimum}
     % \YMdel{the minimum value of the relative}  
     error lower 
     % \YMdel{closer to zero }
     than \YM{that of}
     the conventional method 
     % \YMdel{in Fig.~\ref{fig: result of conventional QA with lambda=0.004}} 
     % \YMdel{regardless of $N$} 
     \YM{for all $N$} (see Table~\ref{table: detail result}). Also, we show that 
     % \YMdel{we can obtain the} 
     the relative error with the RQA-based schedule \YM{can be}
     lower than zero  
     % \YMdel{in Fig.~\ref{fig: result of RQA-based with lambda=0} and \ref{fig: result of RQA-based with lambda=0.004}}
     in some $T$ 
     % \YMdel{regardless of}
     \YM{for all}
     $N$.
     \YM{This means that the RQA-based schedule may produce unphysical results where a corresponding density matrix has negative eigenvalues.}
     % \YMdel{This means that we cannot accurately estimate the ground energy with \YMdel{this} 
     % \YM{RQA-based}
     % schedule by minimizing the expectation value.}
     }
    \label{fig: result of numerical simulation}
  \end{figure*}
In this section, we show the performance of the EMQA schedule and compare it with that of the RQA-based and the conventional schedules. For this purpose, we consider the problem Hamiltonian as \YM{the Heisenberg model described by}
\begin{equation}
H_{\mathrm{P}}=J\sum^{N}_{i=1} (\hat{\sigma}^x_{i}\hat{\sigma}^x_{i+1} + \hat{\sigma}^y_{i}\hat{\sigma}^y_{i+1} + \Delta \hat{\sigma}^z_{i}\hat{\sigma}^z_{i+1}),
\end{equation}
with the periodic boundary condition where $J$ and $\Delta$ are coefficients.
Throughout of this paper,
by setting $J=1$, the time and energy are normalized by this value.
To simplify the discussion, we choose the Lindblad operators as the Pauli matrices $\hat{\sigma}_i^x, \hat{\sigma}_i^y$, and $\hat{\sigma}_i^z$ and assume that the decay rate is constant regardless of the index (i.e. $\lambda_{n}=\lambda$). In our numerical simulation, the GKSL master equation is rewritten by 
% \textcolor{green}{Please remove this.
% \begin{equation}
%     \dfrac{d\rho}{dt} = -i[H(t), \rho]+\dfrac{\lambda}{2}\sum_{i=1}^{N}\sum_{j\in \{x,y,z\}}[\hat{\sigma}_i^j,[\rho, \hat{\sigma}_i^j]].
% \end{equation}
% }
\begin{equation}
    \dfrac{d\rho}{dt} = -i[H(t), \rho]-\dfrac{\lambda}{2}\sum_{i=1}^{N}\sum_{j\in \{x,y,z\}}[\hat{\sigma}_i^j,[\hat{\sigma}_i^j,\rho]].
\end{equation}
Also, we 
% \YMdel{adopt} 
\YM{set} $B/J=1$, $T^{'}J=5$, and $\Delta=-1$.
To evaluate the performances of those schedules, we define a relative error as $(\ev{H_{\mathrm{P}}}-E_{\mathrm{g}}) / \abs{E_{\mathrm{g}}}$ where $\ev{H_{\mathrm{P}}}$ is the expectation value of the problem Hamiltonian obtained with the numerical simulations and $E_{\mathrm{g}}$ denotes the ground-state energy of the problem Hamiltonian.

We plot the relative error against the parameter $T$
\YM{by changing the number of qubits}
% \YMdel{for each number of the qubits $N$ with the numerical simulation}
in Fig.~\ref{fig: result of numerical simulation}. 
% \YMdel{To describe Fig.~\ref{fig: result of numerical simulation}, we set the parameter $T$ as $1$ to $100$ and $\lambda=0, \ 0.004$, and then we simulate QA of those schedules in each case of those parameters $T$ and $\lambda$.} 
\YM{We observe that there is an optimal $T$ to minimize the expectation values. This is because, as we increase $T$, the decoherence becomes more relevant while non-adiabatic transitions become less significant.}
Also, we indicate the minimum values of the expectation values and the relative errors 
% \YMdel{during} 
\YM{with} the numerical simulations in Table~\ref{table: detail result}.
% \YMdel{in each case of the number of the qubits and the schedules when $\lambda=0.004$.}

\begin{table*}
\caption{The minimum value of the expectation value $\ev{H_{\mathrm{P}}}$ (and the relative error) obtained by each schedule with $\lambda=0.004$ and the exact value of the ground-state energy.}
\begin{tabularx}{0.7\textwidth} { 
  | >{\centering\arraybackslash}X 
  | >{\centering\arraybackslash}X 
  | >{\centering\arraybackslash}X
  | >{\centering\arraybackslash}X
  || >{\centering\arraybackslash}X |}
 \hline
 \multicolumn{5}{|c|}{Expectation Value (Relative Error) $\lambda=0.004$} \\
 \hline
 \hline
  & EMQA & RQA-based & Conventional & Exact\\
 \hline
  $N=3$ & $-2.96 \ (0.0119)$ & $-3.41 \ (-0.138)$ & $-2.63 \ (0.124)$ &$-3.00$ \\
\hline
$N=4$ & $-3.95 \ (0.0114)$ & $-5.32 \ (-0.329)$ & $-3.46 \ (0.135)$ &$-4.00$ \\
\hline
$N=5$ & $-4.69 \ (0.0624)$ & $-5.83 \ (-0.167)$ & $-3.84 \ (0.231)$ &$-5.00$ \\
\hline
$N=6$ & $-5.76 \ (0.0395)$ & $-17.2 \ (-1.87)$ & $-4.85 \ (0.191)$ &$-6.00$ \\
\hline
\end{tabularx}
\label{table: detail result}
\end{table*}

% We plot the every line in Fig.~\ref{fig: result of numerical simulation} in each case of the number of the qubits $N=3, 4, 5,$ and $6$.
\YM{In}
Fig.~\ref{fig: result of EMQA with lambda=0} 
% \YMdel{indicates} 
\YM{we show} the result of the numerical simulation with the EMQA schedule 
% \YMdel{when}
\YM{with}
$\lambda=0$. We check 
\YM{that}
the relative errors converge to zero 
\YM{regardless of the number of qubits}
% \YMdel{in the all cases of the number of qubits} 
when there is no decoherence.
% The minimum expected values during the numerical simulation are $-3.00, \ -4.00, \ -5.00$, and $-6.00$, and on the other hand the ground energies are $-3.00, \ -4.00, \ -5.00$, and $-6.00$ when $N=3, \ 4, \ 5,$ and $6$, respectively.
In addition, we show that
\YM{even under the effect of the decoherence with $\lambda/J=0.004$, we can obtain a more accurate estimation of the ground-state energy than the conventional method,}
% \YMdel{this schedule can achieve the higher accuracy of obtaining the expected values than the conventional schedule when $\lambda=0.004$.
% The minimum expected values with the EMQA schedule are closer to the ground energy than  that with the conventional schedule} 
as shown in Table~\ref{table: detail result}.
% When it comes to the relative error, we obtain the minimum values of it as $0.0119, \ 0.0114, \ 0.0624,$ and $0.0395$
% Therefore, we can obtain the closer values to the ground energies by using dual-state purification than that by using the conventional scheme in all cases of the number of qubits.

On the other hand, \YM{in}
Fig.~\ref{fig: result of RQA-based with lambda=0} and \ref{fig: result of RQA-based with lambda=0.004},  
% \YMdel{indicate} 
\YM{we show} the performance of \YM{the method with}
the RQA-based schedule. 
% \YMdel{In the all cases of the number of the qubits, }
\YM{Regardless of the number of qubits,}
the minimum expectation values are lower than the ground energies as shown in Table~\ref{table: detail result}. 
\YM{If we calculate the energy by using ${\rm{Tr}}[\rho H_{\mathrm{P}}]$, 
we always have ${\rm{Tr}}[\rho H_{\mathrm{P}}]\geq E_{\mathrm{g}}$
where $\rho$ denotes an arbitrary density matrix. This means that the method with the RQA-based schedule provides us with an unphysical state.
In this case, even if we minimize the expectation value of the energy by changing $T$, the minimized value is not always the closest to the actual value. This is a significant disadvantage to use the method with the RQA-based schedule.
}
% \YMdel{If we obtain the expected values of observables with physical states in QA, those values are definitely equal to or larger than the true values. Therefore, the minimum expected values obtained by trying several sets of parameters are interpreted as the closest to the true values. However, as we mentioned in the previous section, since we generate the unphysical state in this shcedule, we cannot know which expected value obtained during the QA is the closest to the ground energy.}

Finally, let us discuss the physical implementation of our scheme. 
\YM{It is known that we can use superconducting
flux qubits (or capacitively shunted flux qubits) for both QA and gate-type quantum computation \cite{matsuzaki2020quantum}.
Also, the Hamiltonian of the Heisenberg model with transverse fields can be realized by using these systems \cite{imoto2022obtaining}. For these systems,
the coherence time can be as long as tens of micro seconds \cite{yan2016flux}, and
the coupling strength can be tens of MHz or more \cite{plantenberg2007demonstration}. This means that, by using these quantum devices, we can realize $J/\lambda \simeq 10^3$, which is similar to those used in our simulations. Therefore, these systems are candidates for realizing our proposal.
}

% In the numerical simulation,
% \YS{For the physical implementation of our scheme, }

\section{Conclusion}
\YS{In conclusion,
we propose a QA strategy combined with an EM method to suppress the effects of decoherence. Among many EM methods, we adopt dual-state purification that does not require any two-qubit gates, which is suitable for the devices devised for QA. 
\YM{There are four steps in our protocol. First, let the system evolve by the Hamiltonian. Second, we perform single-qubit projective measurements. Third, let the system evolve by the Hamiltonian whose dynamics corresponds to an inverse map of the first dynamics. Finally, we post-process the measurement results.
We numerically show that, by using our protocol, we can estimate the ground-state energy more accurately than the conventional QA under decoherence.
}
% \YMdel{Though the RQA-based schedule seems to be promising, we show this strategy could bring about the unphysical result in which the expected value is lower than the exact ground energy of the problem Hamiltonian. 
% Meanwhile, we show EMQA does not lead to the unphysical result and provides a more precise estimation than conventional QA.
% Our scheme, EMQA schedule, theoretically gives QA the robustness for decoherence.}
}
% \YMdel{
% In conclusion, we suggest a quantum annealing (QA) schedule named error-mitigated QA (EMQA) to suppress the effects of decoherence with dual-state purification. Dual-state purification does not require two-qubit gates, such as the SWAP gate, or more additional qubits than an original scheme needs, unlike other error mitigation methods like virtual distillation. This point is feasible for QA since implementing the two-qubit gates on QA devices brings about significant technical barriers. In our proposal, we efficiently construct the inverse map this error mitigation method requires.
% We confirm EMQA schedule is superior to the conventional and the reverse QA-based ones in the estimation of the ground energy of the problem Hamiltonian under the decoherence by the numerical simulations. Our method is practical when the problem Hamiltonian contains non-diagonal terms in the computational basis. }
\begin{acknowledgments}
We thank useful advice from S. Endo, K. Yamamoto, H. Hakoshima, and N. Yoshioka.
This work was supported by Leading Initiative for Excellent Young Researchers MEXT Japan and JST presto
(Grant No. JPMJPR1919) Japan. This paper is partly
based on results obtained from a project, JPNP16007,
commissioned by the New Energy and Industrial Technology Development Organization (NEDO), Japan.

We performed the numerical simulations in Fig.~\ref{fig: result of numerical simulation} by using Qutip. This is an efficient framework to simulate quantum mechanics~\cite{johansson2012qutip}.
\end{acknowledgments}

\appendix

\section{Proof for inverse map in EMQA schedule}
\YM{For dual-state purification, we need to construct an inverse transformation of the target unitary dynamics.}
Here, we show \YM{that, with the EMQA schedule,}
the dynamics from $t=T+T^{'}/2$ to $t=2T+T^{'}$ is 
% \YMdel{truly equal} 
\YM{equivalent} to the conjugate transpose of the dynamics from $t=0$ to $t=T+T^{'}/2$ 
% \YMdel{with the EMQA schedule} 
when there is no decoherence. 

First, the dynamics from $t=T+T^{'}/2$ to $t=T+T^{'}$ is equivalent to the conjugate transpose of the dynamics from $t=T$ to $t=T+T^{'}/2$, i.e.,
\begin{eqnarray}
    &&\exp(-i\int_{T+T^{'}/2}^{T+T^{'}}A_t H_{\mathrm{P}}dt)=\exp(iH_{\mathrm{P}}\dfrac{T^{'}}{4}) \nonumber \\
    &=&\exp(-i\int_{T}^{T+T^{'}/2}A_t H_{\mathrm{P}}dt)^{\dag}
    \label{eq: part 1 of proof appendix A}
\end{eqnarray}

% \YMdel{Then} 
\YM{Next}, let us consider the dynamics from $t=T+T^{'}$ to $t=2T+T^{'}$.
% \begin{equation}
%     H_2(t)=\left(\dfrac{t}{T} - \dfrac{T^{'}}{T} - 2\right)H_{\mathrm{P}}+\left(-\dfrac{t}{T} + \dfrac{T^{'}}{T} + 1\right)H_{\mathrm{D}},
% \end{equation}
By transforming the variable $t \ (t:T+T^{'} \rightarrow 2T+T^{'})$ into a variable $s \ (s:0 \rightarrow T)$ by $t=s+T+T^{'}$, we obtain 
\begin{equation}
    H_2(s)=\left(\dfrac{s}{T}-1\right)H_{\mathrm{P}}-\dfrac{s}{T}H_{\mathrm{D}}.
\end{equation}
Thus, \YM{by using the Trotter decomposition}, the dynamics of the Hamiltonian $H_2(s)$ is written as follows: 
\footnotesize
\begin{eqnarray}
    &&\mathcal{T}\exp(-i\int_{0}^{T}H_2(s)ds) \nonumber \\
    &\simeq&\exp\Bigl(-iH_2(M\delta t)\delta t\Bigr)\times\exp\Bigl(-iH_2\bigl((M-1)\delta t\bigr)\delta t\Bigr)\times\cdots \nonumber \\
    &&\times\exp\Bigl(-iH_2(\delta t)\delta t\Bigr)\times\exp\Bigl(-iH_2(0)\delta t\Bigr) \nonumber \\
    &=&\prod_{j=0}^{M}\exp\Bigl(-iH_2(j\delta t)\delta t\Bigr),
\end{eqnarray}
\normalsize
where $\mathcal{T}$ denotes the time-ordered product, 
% \YMdel{the second line represents the Trotter decomposition,} 
$M$ denotes a natural number, and $\delta t=T/M$ denotes a discretized time. Note that $\mathcal{T}\exp(-i\int_{0}^{T}H_2(s)ds)$ is 
% \YMdel{exactly} 
equal to the second line \YM{in the limit of a small $\delta t$}.
% \YMdel{when $\delta t$ is sufficiently small.}
Meanwhile, We define the time-dependent Hamiltonian from $t=0$ to $t=T$ as
\begin{equation}
    H_1(t)=\dfrac{t}{T}H_{\mathrm{P}}+\left(1-\dfrac{t}{T}\right)H_{\mathrm{D}}.
\end{equation}
We can decompose the conjugate transpose of the dynamics from $t=0$ to $t=T$ as 
\footnotesize
\begin{eqnarray}
    &&\Biggl(\mathcal{T}\exp(-i\int_{0}^{T}H_1(t)dt)\Biggr)^{\dag} \nonumber \\
    &\simeq&\Biggl(\exp\Bigl(-iH_1(M\delta t)\delta t\Bigr)\times\exp\Bigl(-iH_1\Bigl((M-1)\delta t\Bigr)\delta t\Bigr)\times\cdots \nonumber \\
    &&\times\exp\Bigl(-iH_1(\delta t)\delta t\Bigr)\times\exp\Bigl(-iH_1(0)\delta t\Bigr) \Biggr)^{\dag}\nonumber \\
    &=&\exp\Bigl(iH_1(0)\delta t\Bigr) \times\exp\Bigl(iH_1(\delta t)\delta t\Bigr) \times\cdots \nonumber \\
    &&\times \exp\Bigl(iH_1\Bigl((M-1)\delta t\Bigr)\delta t\Bigr) \times \exp\Bigl(iH_1(M\delta t)\delta t\Bigr)\nonumber \\
    &=&\prod_{j=0}^{M}\exp\Bigl(iH_1\Bigl((M-j)\delta t\Bigr)\delta t\Bigr).
\end{eqnarray}
\normalsize
% \YMdel{Because} 
\YM{Since} $\exp\Bigl(-iH_2(j\delta t)\delta t\Bigr) = \exp\Bigl(iH_1\bigl((M-j)\delta t\bigr)\delta t\Bigr)$ \YM{is satisfied}
in each index $j$, we have
\footnotesize
\begin{eqnarray}
    &&\mathcal{T}\exp(-i\int_{0}^{T}H_2(s)ds) \nonumber \\
    &=&\lim_{M \to \infty}	
\prod_{j=0}^{M}\exp\Bigl(-iH_2(j\delta t)\delta t\Bigr) 
    =\lim_{M \to \infty}\prod_{j=0}^{M}\exp\Bigl(iH_1\bigl((M-j)\delta t\bigr)\delta t\Bigr) \nonumber \\
    &=&\Biggl(\mathcal{T}\exp(-i\int_{0}^{T}H_1(t)dt)\Biggr)^{\dag}.
    \label{eq: part 2 of proof appendix A}
\end{eqnarray}
\normalsize

% \YMdel{Owing to}
\YM{Therefore, by using}
Eq.~\ref{eq: part 1 of proof appendix A} and \ref{eq: part 2 of proof appendix A}, we show
% \fontsize{8pt}{8pt}\selectfont
\footnotesize
\begin{eqnarray}
    &&\mathcal{T}\exp(-i\int_{T+T^{'}/2}^{2T+T^{'}}\Bigl(A_t H_{\mathrm{P}}+B_t H_{\mathrm{D}}\Bigr)dt) \nonumber \\
    &=&\mathcal{T}\exp(-i\int_{0}^{T}H_2(s)ds)\exp(-i\int_{T+T^{'}/2}^{T+T^{'}}A_t H_{\mathrm{P}}dt)\nonumber \\
    &=&\Biggl(\mathcal{T}\exp(-i\int_{0}^{T}H_1(t)dt)\Biggr)^{\dag}\exp(-i\int_{T}^{T+T^{'}/2}A_t H_{\mathrm{P}}dt)^{\dag}\nonumber \\
    &=&\Biggl(\mathcal{T}\exp(-i\int_{0}^{T+T^{'}/2}\Bigl(A_t H_{\mathrm{P}}+B_t H_{\mathrm{D}}\Bigr)dt)\Biggr)^{\dag}.
\end{eqnarray}
\normalsize
\section{Theoretical analysis for RQA-based schedule}
In the main text, we observe that
% \YMdel{Here, we explain why we obtain} 
the expectation value of $H_{\mathrm{P}}$ \YM{becomes}
lower than the ground energy with the RQA-based method \YM{in the numerical simulations}.
\YM{To obtain a better understanding of this, we
derive analytic expressions of
}
% \YMdel{For this purpose,we analyze the QA of} 
the RQA-based method with a single qubit.
% \YMdel{First,} 
\YM{Actually, we will show that, for this case, the state $\rho\bar{\rho}+\bar{\rho}\rho$ becomes unphysical in the sense that the eigenvalues of this state become lower than 0
% \YMdel{$E_{\mathrm{g}}$}
unlike physical density matrices.}
\YS{In this section, we do not consider the effect of decoherence since we can see such an unphysical situation even if there is no decoherence as shown in Fig.~\ref{fig: result of RQA-based with lambda=0}.}
We set a time-dependent Hamiltonian from $t=0$ to $t=T$ for the QA as
\begin{equation}
    H(t)=-\sin\Biggl(\dfrac{\pi t}{2T}\Biggr)\hat{\sigma}^z -\cos\Biggl(\dfrac{\pi t}{2T}\Biggr)\hat{\sigma}^x.
\end{equation}
We can diagonalize this Hamiltonian with a unitary matrix written as 
% \begin{equation}
% V_t=\dfrac{1}{\sqrt{2}}
%     \begin{pmatrix}
% \sqrt{1-\sin\Bigl(\dfrac{\pi t}{2T}\Bigr)} & -\sqrt{1+\sin\Bigl(\dfrac{\pi t}{2T}\Bigr)} & \ \\
% \ & \ & \ \\
% \sqrt{1+\sin\Bigl(\dfrac{\pi t}{2T}\Bigr)} & \sqrt{1-\sin\Bigl(\dfrac{\pi t}{2T}\Bigr)} & \ \\
% \end{pmatrix}.
% \end{equation}
\begin{equation}
V(t)=
    \begin{pmatrix}
\cos(\dfrac{\pi(t+T)}{4T})& -\sin(\dfrac{\pi(t+T)}{4T})\\
\ & \ \\
\sin(\dfrac{\pi(t+T)}{4T})& \cos(\dfrac{\pi(t+T)}{4T})\\
\end{pmatrix}.
\end{equation}
We obtain a diagonalized matrix as
\begin{equation}
D\equiv
V(t)H(t)V^{\dag}(t)
=
\begin{pmatrix}
1&0 \\
0&-1
\end{pmatrix}.    
\end{equation}
 By using those matrices $V(t)$ and $D$, we define an effective Hamiltonian as 
\begin{eqnarray}
H_{\mathrm{eff}}&=&i \dfrac{dV(t)}{dt}V^{\dag}(t)+D \nonumber \\
    &=&\begin{pmatrix}
1&-\dfrac{i\pi}{4T} \\
\ & \ \\
\dfrac{i\pi}{4T}&-1
\end{pmatrix}.
\end{eqnarray}
% \begin{equation}
% H(t)=\left\{
% \begin{array}{ll}
% \dfrac{t}{T} & (t:0\rightarrow T) \\
% -\dfrac{2t}{T^{'}} + \dfrac{2T}{T^{'}} + 1 & (t:T \rightarrow T + T^{'}) 
% \end{array}
% \right.
% \end{equation}
If there is no decoherence, the time-evolved state $\ket{\psi(T)}$ with the Hamiltonian $H(t)$ from the initial state $\ket{+}$ is written as
\footnotesize
\begin{eqnarray}
    \ket{\psi(T)}&=&\mathcal{T}\exp(-i\int_0^TH(t)dt)\ket{+} \nonumber \\
&=&V^{\dag}(T)\exp(-iTH_{\mathrm{eff}})V(0)\ket{+} \nonumber \\
&=&\left(
\begin{array}{c}
\cos\Biggl(\dfrac{1}{4}\sqrt{\pi^2+16T^2}\Biggr)+\dfrac{4iT\sin\Biggl(\dfrac{1}{4}\sqrt{\pi^2+16T^2}\Biggr)}{\sqrt{\pi^2+16T^2}} \\
\\
\dfrac{\pi\sin\Biggl(\dfrac{1}{4}\sqrt{\pi^2+16T^2}\Biggr)}{\sqrt{\pi^2+16T^2}}
\end{array}
\right). \nonumber \\
\end{eqnarray}
\normalsize
In this case, we obtain the density matrix $\rho$ as $\ketbra*{\psi(T)}$.

\begin{figure}[h!]
  \includegraphics[width=0.45\textwidth]{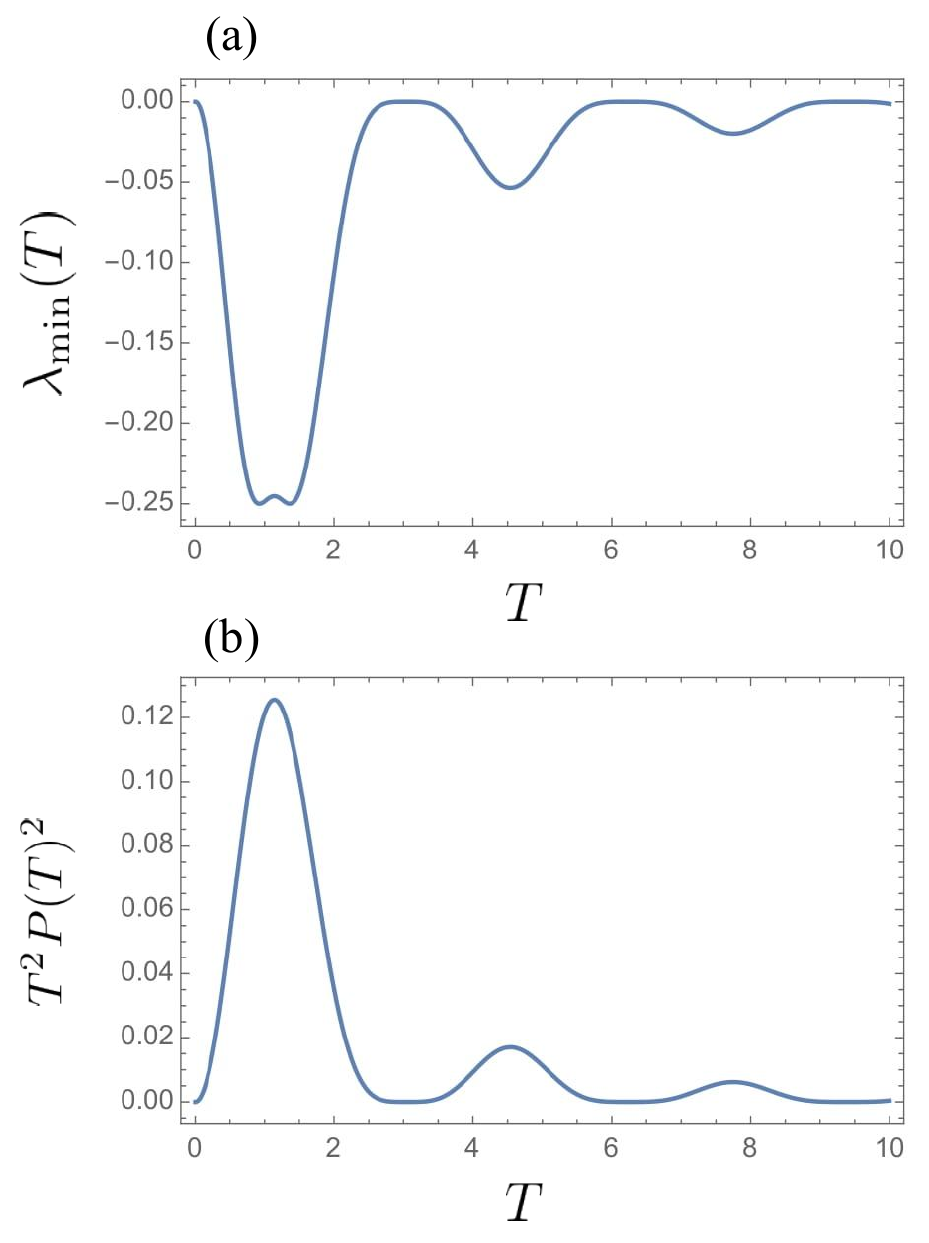}
  \caption{The relationship between the transition rate $P(T)$ in Eq.~\eqref{eq: transition rate} and the minimum eigenvalue $\lambda_{\mathrm{min}}(T)$ in Eq.~\eqref{eq: definition of lambda min} of the state $\rho\bar{\rho}+\bar{\rho}\rho$. We plot $\lambda_{\mathrm{min}}(T)$ and $T^2P(T)^2$ against the annealing time $T$ in Fig.~\ref{fig: lambda and transition rate in Appendix B} (a) and (b), respectively. We check they have peaks at the same points of $T$. This is the evidence in which the eigenvalue is deeply connected with the non-adiabatic transition as shown by Eq.~\eqref{eq: conclusion of Appendix B}.}
    \label{fig: lambda and transition rate in Appendix B}
\end{figure}

On the other hand, the time-dependent Hamiltonian from $t=T$ to $t=2T$ which corresponds to the inverse map $\mathcal{G}$ is defined as
\begin{equation}
    H(t)=-\cos\Biggl(\dfrac{\pi}{2T}(t-T)\Biggr)\hat{\sigma}^z -\sin\Biggl(\dfrac{\pi}{2T}(t-T)\Biggr)\hat{\sigma}^x.
    \label{eq: latter Hamiltonian in Appendix B}
\end{equation}
\YS{
We can also diagonalize this Hamiltonian with a unitary matrix written as
\begin{equation}
U(t)=
    \begin{pmatrix}
\sin(\dfrac{\pi(t-T)}{4T})& -\cos(\dfrac{\pi(t-T)}{4T})\\
\ & \ \\
\cos(\dfrac{\pi(t-T)}{4T})& \sin(\dfrac{\pi(t-T)}{4T})\\
\end{pmatrix}.
\end{equation}
% By using this matrix $U(t)$, 
We obtain an effective Hamiltonian during this time range as 
\begin{eqnarray}
H_{\mathrm{eff}}^{'}&=&i \dfrac{dU(t)}{dt}U(t)^{\dag}+D \nonumber \\
    &=&\begin{pmatrix}
1&\dfrac{i\pi}{4T} \\
\ & \ \\
-\dfrac{i\pi}{4T}&-1
\end{pmatrix}
\end{eqnarray}
because the diagonalized matrix of the Hamiltonian in Eq.~\eqref{eq: latter Hamiltonian in Appendix B} is also $D$.
}
By using this effective Hamiltonian, we define a state $\ket{\phi(T)}$ as
\footnotesize
\begin{eqnarray}
    \ket{\phi(T)}&=& \Biggl(\mathcal{T}\exp(-i\int_{T}^{2T}H(t)dt)\Biggr)^{\dag}\ket{+}\nonumber \\
&=& \Biggl(U(2T)^{\dag}\exp(-iTH_{\mathrm{eff}}^{'})U(T)\Biggr)^{\dag}\ket{+}\nonumber \\
&=&\left(
\begin{array}{c}
\cos\Biggl(\dfrac{1}{4}\sqrt{\pi^2+16T^2}\Biggr)-\dfrac{4iT\sin\Biggl(\dfrac{1}{4}\sqrt{\pi^2+16T^2}\Biggr)}{\sqrt{\pi^2+16T^2}} \\
\\
\dfrac{\pi\sin\Biggl(\dfrac{1}{4}\sqrt{\pi^2+16T^2}\Biggr)}{\sqrt{\pi^2+16T^2}}
\end{array}
\right).\nonumber\\
\end{eqnarray}
\normalsize
By using this state, we have $\bar{\rho}$ in Eq.~\ref{eq: dual-state purification} as $\ketbra*{\phi(T)}$.
% \YMdel{here we only consider} 
\YM{For our purpose, 
it is sufficient to consider only}
$\rho\bar{\rho}+\bar{\rho}\rho$
because
% \YM{Since} 
$\Tr\Bigl[\rho\bar{\rho}+\bar{\rho}\rho\Bigr]>0$ is satisfied in this case.
We obtain the minimum eigenvalue $\lambda_{\mathrm{min}}(T)$ of the state $\rho\bar{\rho}+\bar{\rho}\rho$ as 
\begin{equation}
    \lambda_{\mathrm{min}}(T)=\dfrac{A-(\pi^2+16T^2)\sqrt{A}}{(\pi^2+16T^2)^2},
    \label{eq: definition of lambda min}
\end{equation}
where
\begin{eqnarray}
    A=&&\pi^4+8\pi^2T^2+256T^4+32\pi^2T^2\cos\Biggl(\dfrac{1}{2}\sqrt{\pi^2+16T^2}\Biggr) \nonumber \\
    &&-8\pi^2T^2\cos\Biggl(\sqrt{\pi^2+16T^2}\Biggr).
\end{eqnarray}
We plot $\lambda_{\mathrm{min}}(T)$  against the parameter $T$ in Fig.~\ref{fig: lambda and transition rate in Appendix B} (a).
As we can see in Fig.~\ref{fig: lambda and transition rate in Appendix B} (a), $\lambda_{\mathrm{min}}(T)$ can be negative 
% \YMdel{in some cases of} 
\YM{for some} $T$.
Thus, the state $\rho\bar{\rho}+\bar{\rho}\rho$ is unphysical 
\YM{in the sense that the eigenvalue is negative}
even if we normalize this state. 
\YM{As long as the state $\rho\bar{\rho}+\bar{\rho}\rho$ has a negative eigenvalue, 
the expectation values of the energy
$\Tr\Bigl[H_{\mathrm{P}}(\rho\bar{\rho}+\bar{\rho}\rho)\Bigr]/\Tr\Bigl[\rho\bar{\rho}+\bar{\rho}\rho\Bigr]$ can be lower than the ground-state energy.}
% \YMdel{
% This unphysical state can lead to the expected value $\Tr\Bigl[H_{\mathrm{P}}(\rho\bar{\rho}+\bar{\rho}\rho)\Bigr]/\Tr\Bigl[\rho\bar{\rho}+\bar{\rho}\rho\Bigr]$ becoming lower than the ground energy of $H_{\mathrm{P}}$ by changing $T$.
% }

\YM{We expect that this negative eigenvalue of $\rho\bar{\rho}+\bar{\rho}\rho$ is related to the non-adiabatic transitions during QA.
\YS{Actually, we confirm that the unphysical state appears at relatively small $T$ as shown in Fig.~\ref{fig: result of RQA-based with lambda=0}.}
For further investigation,}
% \YMdel{To examine the relationship between $\lambda_{\mathrm{min}}(T)$ and the non-adiabatic transition,} 
we define a transition rate $P(T)$ as
\begin{equation}
    P(T)=|\braket{1}{\psi(T)}|^2 = \dfrac{\pi^2 \sin^2\Biggl(\dfrac{\sqrt{\pi^2+16T^2}}{4}\Biggr)}{\pi^2+16T^2}.
    \label{eq: transition rate}
\end{equation}
% \YMdel{If $P(T)=0$, this means the non-adiabatic transition does not occur.}
\YM{When $P(T)$ is not zero, there are non-adiabatic transitions during the dynamics.}
By using the transition rate $P(T)$, we rewrite
\begin{equation}
    \lambda_{\mathrm{min}}(T)=\Biggl(\sqrt{1-\dfrac{64T^2P(T)^2}{\pi^2}}-\dfrac{1}{2}\Biggr)^2-\dfrac{1}{4}.
    \label{eq: conclusion of Appendix B}
\end{equation}
% We draw the relationship between $\lambda_{\mathrm{min}}(T)$ and $T^2 P(T)^2$ in Fig.~\ref{fig: lambda and transition rate in Appendix B}. 
We plot $T^2 P(T)^2$ against the parameter $T$ in Fig.~\ref{fig: lambda and transition rate in Appendix B} (b).
% We plot $T^2 P(T)^2$ and $\lambda_{\mathrm{min}}(T)$  against the parameter $T$ in Fig.~\ref{fig: lambda and transition rate in Appendix B} (a) and (b), respectively.
We confirm that the eigenvalue $\lambda_{\mathrm{min}}(T)$ decreases (increases) as $T^2 P(T)^2$ increases (decreases) in Fig.~\ref{fig: lambda and transition rate in Appendix B}.
%Thus, 
\YM{This shows that the non-adiabatic transitions are deeply related to the negative eigenvalues of the state $\rho\bar{\rho}+\bar{\rho}\rho$ in Eq.~\eqref{eq: conclusion of Appendix B}.}
% \YMdel{we conclude the negative eigenvalue derives from the non-adiabatic transition.}

\bibliographystyle{apsrev4-1}
\bibliography{emqa}
\end{document}